\documentclass[a4paper,11pt]{article}
\pdfoutput=1

\usepackage{jheppub}
\usepackage[T1]{fontenc}
\usepackage{lmodern}
\usepackage{microtype}
\usepackage{array}
\usepackage{booktabs}
\usepackage[section]{placeins}
\usepackage{longtable}
\usepackage{tabularx}
\usepackage{tikz-cd}
\usetikzlibrary{decorations.pathmorphing}
\usepackage{siunitx}
\title{SMEFT Constraints on Top Electroweak Couplings and a Composite Higgs Interpretation}

\author{Leandro Da Rold}
\author{and Juan Fernandez}
\affiliation{Centro At\'omico Bariloche, Instituto Balseiro and CONICET,\\
Av.~Bustillo 9500, 8400 S.~C.~de Bariloche, Argentina}
\emailAdd{daroldl@ib.edu.ar}
\emailAdd{juan.fernandez@ib.edu.ar}

\abstract{
Recent LHC measurements of $tWZ$, $t\bar t Z$ and single top production provide complementary probes of top quark electroweak interactions. We incorporate these measurements into a global Standard Model effective field theory analysis of 46 independent dimension-six Wilson coefficients at linear order. The updated fit remains compatible with the Standard Model at $95\%$ CL, and favors negative values of $C_{Hq}^{(-)}\equiv C_{Hq}^{(1)}-C_{Hq}^{(3)}$ and $C_{Ht}$, with central values suggestive of TeV scale new physics. We also present an effective composite Higgs model based on a two-site description that realizes this pattern of Wilson coefficients through partial compositeness. The model includes custodial symmetry and protection of the $Z\bar b_L b_L$ coupling. A fermion bitriplet generates the required left-handed shift, and an additional embedding of the right-handed top allows negative contributions to $C_{Ht}$ from both fermion and spin-one exchange. Scanning four right-handed top embeddings, we find parameter regions within the $68\%$ CL contour that are compatible with electroweak precision data, Higgs coupling measurements and applicable direct searches. The $T$ parameter limits the negative fermionic contribution to $C_{Ht}$, favoring a spin-one contribution of comparable or greater magnitude. Larger $SU(2)_R$ representations extend the reach toward negative $C_{Ht}$, and also enhance four-quark interactions that increase the four-top cross section, in the direction favored by current measurements. The mechanism connects the preferred top electroweak shifts to complementary tests through single production of quarks with electric charge $-4/3$, modified Higgs couplings and four-top production. }

\arxivnumber{2611.05500}

\begin{document}
\maketitle
\flushbottom

\section{Introduction}
The Standard Model (SM) of elementary particles has been tested by a vast number of experiments over the past decades, meeting every challenge with impressive precision~\cite{ParticleDataGroup:2024cfk}. Yet many fundamental questions remain open, such as the stability of the electroweak scale~\cite{Susskind:1978ms}, the origin of the matter--antimatter asymmetry of the Universe~\cite{Sakharov:1967dj}, the existence of dark matter~\cite{Planck:2018vyg} and the origin of neutrino masses~\cite{Super-Kamiokande:1998kpq}. The answers to some of these questions may lie at the electroweak scale, the realm of the electroweak gauge bosons, the Higgs boson and the top quark. The LHC, which is probing this scale up to a few TeV, will be able to study the properties of these particles with unprecedented precision, as well as new phenomena associated with them~\cite{ATLAS:2025eii}.

ATLAS and CMS have direct access to the couplings of the top quark with the electroweak gauge bosons through measurements of single and double top final states in association with electroweak gauge bosons. These measurements are highly sensitive to deviations from the SM, since beyond the SM (BSM) states can generate anomalous electroweak top couplings~\cite{Miralles:2021dyw}. Recently, CMS reported the first observation of $tWZ$ production in proton-proton collisions using $200~\mathrm{fb}^{-1}$ of Run~2 and Run~3 data~\cite{CMS:2025tre}. The signal was established with an observed (expected) significance of $5.8\sigma$ ($3.5\sigma$). Defining the signal strength as $\mu_X\equiv\sigma_X/\sigma_X^{\mathrm{SM}}$ for a process $X$, the nominal fit gives
\begin{equation}
  \mu_{tWZ} = 1.77\pm0.23~(\mathrm{stat.})\pm0.22~(\mathrm{syst.}),
\end{equation}
which lies $2.3\sigma$ above the SM prediction.

Since $t\bar t Z$ is the dominant background and overlaps with $tWZ$ beyond leading order, CMS also performed a simultaneous fit without an external constraint on $t\bar t Z$, which gives
\begin{equation}
  \mu_{tWZ}=1.88\pm0.34,
  \qquad
  \mu_{t\bar t Z}=0.91\pm0.05,
\end{equation}
where the uncertainties have been combined in quadrature.

ATLAS has presented the full Run~2 results on the production of $t\bar t Z$, including the differential distributions~\cite{ATLAS:2023eld}. An interpretation of these measurements in terms of an effective field theory shows a pattern compatible with the CMS results.

In this article, first we explore whether the measured top electroweak couplings could be a hint of new physics (NP). Deviations from the SM, if the NP scale $\Lambda$ lies above the energies directly probed by the experiments, can be described by an effective field theory. The SM effective field theory (SMEFT) provides a general and predictive framework to study precision measurements and to interpret deviations in terms of physics beyond the SM~\cite{Buchmuller:1985jz,Grzadkowski:2010es}. However, since the number of independent operators is already large at dimension six, given the limited number of observables and the degeneracies in the operator space, the interpretation of the NP in terms of the SMEFT is not straightforward, and identifying a pattern of deviations can provide a clue. Restricting to the operators that modify the top electroweak couplings at leading order in $1/\Lambda$, we show that the measurements of $tWZ$, $t\bar tZ$ and $b\bar bZ$ favor a pattern of Wilson coefficients (WCs): $C_{Hq}^{(-)}<0$, $C_{Ht}<0$ and $C_{Hq}^{(+)}\simeq 0$. Imposing the flavor symmetry $U(2)_q\times U(2)_u\times U(3)_d\times[U(1)_\ell\times U(1)_e]^3$~\cite{DAmbrosio:2002vsn,Degrande:2020evl}, which distinguishes the top quark from the light fermions, on the full set of SMEFT WCs, and using SMEFiT~\cite{Giani:2023gfq}, we show that this pattern persists after the inclusion of the most recent measurements sensitive to these couplings.

Having identified a pattern of WCs that hints at NP in the top electroweak sector, we investigate a UV theory that could match it. We find that no single new particle can account for this pattern. Instead, we consider a composite Higgs scenario with spin-one and fermionic resonances that mix linearly with the elementary SM fields~\cite{Kaplan:1991dc,Contino:2006nn}. To avoid excessive contributions to precisely measured electroweak observables, the composite sector is protected by custodial symmetry $SU(2)_L\times SU(2)_R$~\cite{Agashe:2003zs}, and the mixings satisfy a discrete symmetry that protects the $Z\bar b_L b_L$ coupling~\cite{Agashe:2006at}. We find a set of fermionic resonances that generates the top quark Yukawa and matches the pattern obtained in the fit. The minimal setup turns out to require a bidoublet $Q\sim(\mathbf 2,\mathbf 2)$ and a singlet $T_1\sim(\mathbf 1,\mathbf 1)$ for the top Yukawa, together with a bitriplet of fermions $\Xi\sim(\mathbf 3,\mathbf 3)$ coupled to $Q$, an $SU(2)_R$ multiplet $T_n\sim(\mathbf 1,\mathbf n)$ mixing with $t_R$, and a $Q'_n\sim(\mathbf 2,\mathbf{n-1})$ with Yukawa couplings to $T_n$. We show that the new fermions carry the quantum numbers required to generate the preferred pattern of deviations in the top electroweak couplings.

As is well known, the new spin-one resonances generate tree level contributions to the $S$ parameter~\cite{Agashe:2004rs,Agashe:2005dk,Contino:2006nn}. Despite custodial protection, the composite fermions induce a contribution to the $T$ parameter at the one-loop level, as well as one-loop effects in $S$~\cite{Agashe:2003zs,Carena:2007ua}. We scan the parameter space of the composite Higgs model (CHM) and determine the region that reproduces the pattern of WCs while satisfying the electroweak precision constraints. We find that, since in the model there is a correlation between $T$ and the fermionic contributions to $C_{Ht}$, the constraints from $T$ limit the fermionic contribution to $C_{Ht}$, favoring a spin-one contribution of comparable or larger magnitude in the region selected by the fit.

A distinctive signature of the model is the presence of fermionic resonances with charge $-4/3$, which could be discovered in single production at the High Luminosity LHC, as well as states with charges $5/3$ and $8/3$~\cite{CMS:2024bni,Matsedonskyi:2014lla}. It also predicts deviations in the top and bottom quark Yukawa couplings, in the Higgs couplings to photons and gluons, as well as top WCs that could be tested in four-top production~\cite{Pomarol:2008bh} and $t\bar tb\bar b$~\cite{Aoude:2022deh,DHondt:2018cww}. An extension of the linear fit constrains two combinations of these coefficients, both compatible with the model predictions.

The article is organized as follows: in sec.~\ref{sec:smeft} we present an analysis of $tWZ$ and $t\bar t Z$ in the SMEFT, as well as a global fit of the WCs including the available data on top quark measurements from the Tevatron and the LHC. In sec.~\ref{sec:2site} we present a CHM with resonances, in terms of a two-site model, and we show the WCs of the SMEFT as well as the contributions to electroweak observables. We show the numerical predictions of the model in sec.~\ref{sec:num} and its characteristic signals in sec.~\ref{sec:sig}. We conclude in sec.~\ref{sec:conc} and show some details of the calculations in the appendices.

\section{Top quark data and global SMEFT analysis}
\label{sec:smeft}
In this section we consider measurements of the top electroweak couplings in the context of the SMEFT. First we consider only the operators that involve the top quark and the electroweak gauge bosons, and show the directions preferred by the recent data. After that we perform a global fit of the WCs with SMEFiT, adding the measurements in table~\ref{tab:data}.

\subsection{Top electroweak measurements and preferred SMEFT directions}
\label{sec:twz}
We restrict the discussion of this section to the operators built from third-generation quark fields and bosons whose interference with the SM amplitude is unsuppressed. Six such operators of the Warsaw basis~\cite{Grzadkowski:2010es} modify the $tWZ$ and $t\bar t Z$ rates at linear order,
\begin{align}
  \mathcal O_{Hq}^{(1)}
    &=i(H^\dagger\overleftrightarrow{D}_\mu H)(\bar q\gamma^\mu q),
  &
  \mathcal O_{tW}
    &=(\bar q\sigma^{\mu\nu}t)\,\tau^a\widetilde H\,W_{\mu\nu}^a,
  \notag\\
  \mathcal O_{Hq}^{(3)}
    &=i(H^\dagger\overleftrightarrow{D}_\mu\tau^a H)(\bar q\tau^a\gamma^\mu q),
  &
  \mathcal O_{tB}
    &=(\bar q\sigma^{\mu\nu}t)\,\widetilde H\,B_{\mu\nu},
  \notag\\
  \mathcal O_{Ht}
    &=i(H^\dagger\overleftrightarrow{D}_\mu H)(\bar t\gamma^\mu t),
  &
  \mathcal O_{tG}
    &=g_s(\bar q\sigma^{\mu\nu}T^At)\,\widetilde H\,G_{\mu\nu}^A,
  \label{eq:ops}
\end{align}
with $\sigma^{\mu\nu}=\frac i2[\gamma^\mu,\gamma^\nu]$ and Hermitian conjugates understood for the dipoles, in the conventions of Ref.~\cite{Celada:2024mcf}. The left column collects the current operators and the right column the dipoles.

We write the linear response of each cross section as
\begin{equation}
  \frac{\sigma_X}{\sigma_X^{\rm SM}}
  =1+\sum_i a_i^X C_i+\mathcal O(C_i C_j),
  \qquad X=tWZ,\,t\bar t Z,
  \label{eq:mu}
\end{equation}
where $a_i^X$ has units of $\mathrm{TeV}^2$ when $C_i$ is expressed in $\mathrm{TeV}^{-2}$.

Table~\ref{tab:interf} shows the response of the two rates to one operator at a time. Since the two columns correspond to different process definitions, the comparison is restricted to signs and relative sensitivities. Only three directions enhance $tWZ$ appreciably: $C_{Hq}^{(3)}>0$, $C_{tW}<0$ and $C_{tG}>0$. The two dipoles are severely constrained by other measurements. The constraint on $C_{tG}$ comes mainly from inclusive $t\bar t$ production~\cite{Celada:2024mcf}, and $C_{tG}>0$ is further disfavored by the measured $t\bar tZ$ rate, which it enhances much more strongly than $tWZ$. The constraint on $C_{tW}$ comes mainly from $W$ helicity fractions and from other analyses~\cite{Cornet-Gomez:2025jot,Miralles:2021dyw}, which exclude a negative $C_{tW}$ large enough to produce the $tWZ$ excess. The remaining direction is $C_{Hq}^{(3)}>0$, which also enhances $t\bar tZ$; since that rate is measured far more precisely than $tWZ$ and lies below the SM, a negative $C_{Ht}$ is needed to compensate, and it does so with a much smaller effect on $tWZ$. The two correlated rates therefore select the current directions $C_{Hq}^{(3)}>0$ and $C_{Ht}<0$, probed most directly in associated production of a top quark with an electroweak boson.
\begin{table}[t]
  \centering
  \small
  \renewcommand{\arraystretch}{1.13}

  \begin{tabular*}{0.62\textwidth}{
    @{\extracolsep{\fill}} l
    S[table-format=+2.1,detect-weight=true,retain-explicit-plus]
    S[table-format=+2.1,detect-weight=true,retain-explicit-plus] @{}
  }
    \toprule
    \multicolumn{1}{c}{Coefficient}
      & \multicolumn{1}{c}{$tWZ$ [\%]}
      & \multicolumn{1}{c}{$t\bar t Z$ [\%]} \\
    \midrule
    $C_{Hq}^{(1)}$ & +2.6  & -10.4 \\
    \bfseries $C_{Hq}^{(3)}$ & \bfseries +17.4 & \bfseries +10.4 \\
    $C_{Ht}$       & +1.7  & +6.6  \\
    \addlinespace[2pt]
    \bfseries $C_{tW}$       & \bfseries -11.2 & \bfseries -0.3  \\
    $C_{tB}$       & +0.2  & +0.2  \\
    \bfseries $C_{tG}$       & \bfseries +6.8  & \bfseries +35.5 \\
    \bottomrule
  \end{tabular*}
  \caption{Interference responses at $\sqrt{s}=13~\mathrm{TeV}$, computed at NLO in QCD. The entries are $100\,a_i^X$ per $\mathrm{TeV}^{-2}$, with $a_i^X$ defined in Eq.~\eqref{eq:mu}. For $tWZ$ we show the DR2 prediction with the $b$-jet veto of Ref.~\cite{Faham:2021zet}, translated to our coefficient convention, and for $t\bar tZ$ the inclusive on shell prediction of Ref.~\cite{BessidskaiaBylund:2016jvp}, converted to the same normalization.}
  \label{tab:interf}
\end{table}

The SMEFT interpretation of the ATLAS full Run~2 $t\bar t Z$ differential distributions shows a compatible pattern~\cite{ATLAS:2023eld}. In the marginalized linear fit $C_{Hq}^{(3)}$ prefers positive values and $C_{Hq}^{(1)}$ negative ones, both compatible with the SM, while the $95\%$ credible interval for $C_{Ht}$ prefers a negative value while excluding zero.

At tree level the three current operators of Eq.~\eqref{eq:ops} shift four electroweak vertices of the third generation, the neutral $Z\bar b_L b_L$, $Zt_L\bar t_L$ and $Zt_R\bar t_R$ and the charged $Wt_L\bar b_L$~\cite{Celada:2024mcf}. Writing these interactions as
\begin{equation}
{\mathcal L}\supset(g/c_W)Z_\mu\,\bar f\gamma^\mu(g_{Zf_L}P_L+g_{Zf_R}P_R)f
+(g/\sqrt2)\,g_{Wtb}W_\mu^+\,\bar t\gamma^\mu P_Lb+{\rm h.c.},
\end{equation}
with $c_W=\cos\theta_W$ and $g_{Wtb}=1$ in the SM, it is standard to write the doublet operators as $C_{Hq}^{(\pm)}\equiv C_{Hq}^{(1)}\pm C_{Hq}^{(3)}$,
\begin{equation}
  \begin{aligned}
    \delta g_{Zb_L}&=-\frac{v^2}{2}C_{Hq}^{(+)},
    &\qquad
    \delta g_{Zt_L}&=-\frac{v^2}{2}C_{Hq}^{(-)},
    \\[2pt]
    \delta g_{Zt_R}&=-\frac{v^2}{2}C_{Ht},
    &\qquad
    \delta g_{Wtb}&=\frac{v^2}{2}\left(C_{Hq}^{(+)}-C_{Hq}^{(-)}\right),
  \end{aligned}
\end{equation}
where $v=(\sqrt2G_F)^{-1/2}\simeq246~\mathrm{GeV}$ is the electroweak scale. The shift in $g_{Wtb}$ is not independent, $\delta g_{Wtb}=\delta g_{Zt_L}-\delta g_{Zb_L}$.

LEP and SLC constrain $\delta g_{Zb_L}$ tightly~\cite{ALEPH:2005ab}, so $C_{Hq}^{(+)}\simeq0$ and the remaining deviations, $\delta g_{Zt_R}$ and $\delta g_{Zt_L}\simeq\delta g_{Wtb}$, are governed by $C_{Hq}^{(-)}$ and $C_{Ht}$. The preference found above then reads $C_{Hq}^{(-)}<0$ and $C_{Ht}<0$, with all three couplings shifted upward. The implications of these measurements must be assessed together with the other observables sensitive to the same top electroweak interactions, which requires a global analysis.

\subsection{Global fit setup and results}
\label{sec:fit}

We perform a frequentist global analysis at linear order in the Wilson coefficients specified in Appendix~\ref{app:coords}. The $\chi^2$ is constructed with SMEFiT v3.1.1 \cite{Giani:2023gfq}, following the conventions of the published SMEFiT3.0 analysis~\cite{Celada:2024mcf}, and, being quadratic in the WCs, is minimized and profiled exactly. The public v3.1 database \cite{SMEFiTDatabase:v3.1} defines our reference dataset, and the updated one includes the measurements in table~\ref{tab:data}, which replace earlier measurements of the same processes. Measurements with substantially overlapping event samples are not used simultaneously. Both fits include NLO QCD corrections whenever available, and the remaining setup is common to the two and is specified in Appendix~\ref{app:fit}. The one-dimensional profiles for all $46$ fitted Wilson coefficients are collected in Appendix~\ref{app:prof}.

\begin{table}[htbp]
  \centering
  \setlength{\tabcolsep}{5pt}
  \renewcommand{\arraystretch}{1.15}
  \begin{tabularx}{0.93\textwidth}{
    @{}
    l
    l
    c
    c
    >{\raggedright\arraybackslash}X
    c
    @{}
  }
    \toprule
    Process
      & Experiment
      & $\sqrt{s}$ [TeV]
      & $\mathcal{L}$ [fb$^{-1}$]
      & Fit information
      & Reference \\
    \midrule
    \multicolumn{6}{@{}l}{\textit{Measurements sensitive to top electroweak couplings}} \\
    \addlinespace[2pt]
    $tWZ,\ t\bar t Z$
      & CMS
      & $13,\ 13.6$
      & $200$
      & Correlated signal strengths
      & \cite{CMS:2025tre} \\
    $t\bar t Z$
      & ATLAS
      & $13$
      & $140$
      & Fisher compressed SMEFT information
      & \cite{ATLAS:2023eld} \\
    $tW$
      & ATLAS
      & $13$
      & $140$
      & Inclusive cross section
      & \cite{ATLAS:2024ppp} \\
    $tW$
      & CMS
      & $13$
      & $138$
      & Inclusive cross section
      & \cite{CMS:2022ytw} \\
    $tq,\ \bar t q$
      & ATLAS
      & $13$
      & $140$
      & Reduced EFT fit
      & \cite{ATLAS:2025adk} \\
    \midrule
    \multicolumn{6}{@{}l}{\textit{Other added top quark inputs}} \\
    \addlinespace[2pt]
    $t\bar t W^\pm$
      & ATLAS
      & $13$
      & $140$
      & Inclusive cross section
      & \cite{ATLAS:2024moy} \\
    $t\bar t W^\pm$
      & CMS
      & $13$
      & $138$
      & Inclusive cross section
      & \cite{CMS:2022tkv} \\
    $t\bar t$ $A_{\mathrm{FB}}$
      & CDF${}+{}$D0
      & $1.96$
      & $\simeq 10$
      & Combined inclusive forward-backward asymmetry
      & \cite{CDF:2017cvy} \\
    \bottomrule
  \end{tabularx}
  \caption{Recent top quark measurements included in the updated dataset.}
  \label{tab:data}
\end{table}

For $t\bar t Z$ we do not use the Bayesian intervals quoted above but the Fisher compressed information of the unfolded distributions, which enters the fit like the other measurements. The correlation between the CMS $tWZ$ and $t\bar t Z$ rates is incorporated through a Gaussian approximation to the published two-dimensional likelihood scan~\cite{CMS:2025tre}, rather than by treating the two rates as independent measurements. The recent ATLAS measurement of $tZq$~\cite{ATLAS:2026cwj} shows the same preference for a negative $C_{Ht}$ and a negative $C_{Hq}^{(-)}$. Since the $tZq$ measurements share events with the $t\bar tZ$ and $tWZ$ measurements of the same experiment, we do not include them.

The ATLAS and CMS $tW$ cross sections constrain $C_{Hq}^{(3)}$ through the left-handed $Wtb$ vertex. With $C_{Hq}^{(+)}$ fixed by the $Z$-pole data this bounds $C_{Hq}^{(-)}$ without involving the neutral current. The ATLAS $t$-channel single top analysis is sensitive to both electroweak current and four-fermion coefficients. The $t\bar t W^\pm$ measurements and the Tevatron forward-backward asymmetry have little direct sensitivity to the three coefficients of interest and mostly constrain four-fermion directions varied in the fit. For the Tevatron asymmetry we use the LO QCD SMEFT dependence of Ref.~\cite{Rosello:2015sck}. These inputs therefore reduce degeneracies in the $(C_{Hq}^{(-)},C_{Ht})$ plane.

To obtain joint constraints on $(C_{Hq}^{(-)},C_{Ht})$, we scan this plane and profile the remaining $44$ independent coefficients, including the orthogonal combination $C_{Hq}^{(+)}$. The confidence regions follow from $\Delta\chi^2_{\mathrm{prof}}=\chi^2_{\mathrm{prof}}-\chi^2_{\min}$ with the standard two-parameter thresholds~\cite{Cowan:2010js}.

\begin{figure}[!htbp]
  \centering
  \includegraphics[width=0.94\textwidth]
    {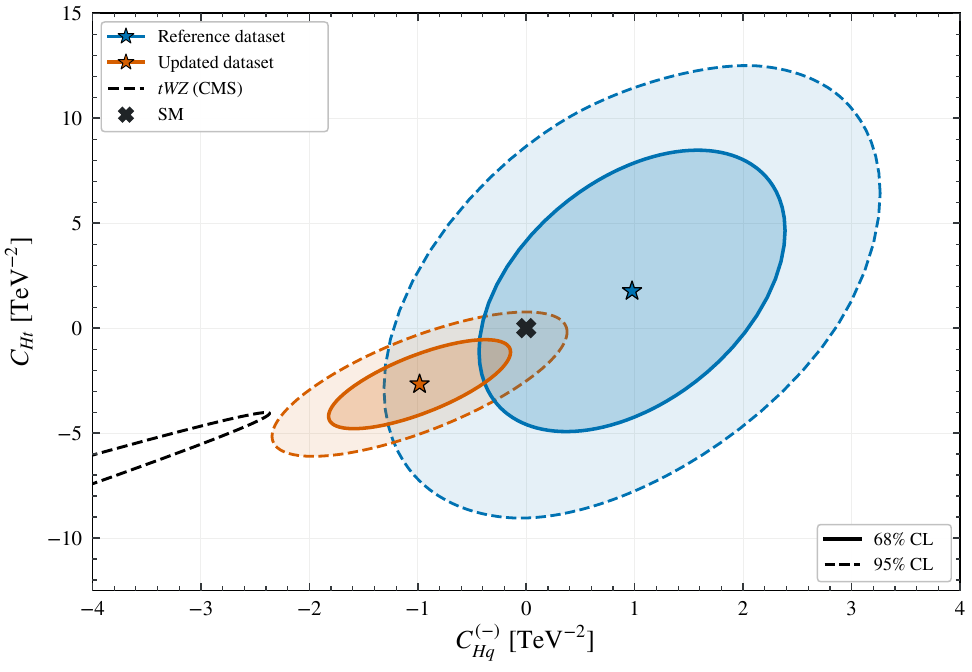}
  \caption{Profiled confidence regions in the $(C_{Hq}^{(-)},C_{Ht})$ plane for the reference (SMEFiT v3.1) and updated datasets. Solid and dashed contours denote the $68\%$ and $95\%$ confidence levels, respectively. The black contour is the $95\%$ confidence region obtained from the CMS $tWZ$ likelihood~\cite{CMS:2025tre} alone, with all other coefficients set to zero.}
  \label{fig:fit}
\end{figure}

Relative to the reference fit, the updated fit contracts the allowed region and shifts the best fit point from $C_{Hq}^{(-)}>0$ and $C_{Ht}>0$ toward $C_{Hq}^{(-)}<0$ and $C_{Ht}<0$, the direction selected by the correlated rates of Sec.~\ref{sec:twz}, as shown in Fig.~\ref{fig:fit}. The SM point lies within the $68\%$ confidence region of the reference fit, but after including the new data it lies at approximately $86\%$ confidence level, outside the updated $68\%$ contour. All but two of the one-dimensional $95\%$ confidence intervals of Appendix~\ref{app:prof} include the SM value, and the exceptions are discussed there.

The same figure shows the $95\%$ confidence region obtained from the CMS $tWZ$ likelihood alone. It is still broad, so its impact on the global fit is limited and the direction is selected only in combination with the other top measurements. Improved measurements with the full Run~3 dataset and at the HL-LHC will test whether the tendency persists.

\section{A custodial two-site completion}
\label{sec:2site}
In the previous section we used top quark data to constrain the SMEFT coefficients of the top electroweak sector. In this section we consider an extension of the SM able to generate the preferred pattern of deviations. Since these are confined to the couplings of the top quark, we look for a new sector coupled mainly to the third generation.

\subsection{From the SMEFT fit to a custodial sector}

The global fit favors the third-generation electroweak current direction
\begin{equation}
  C_{Hq}^{(-)}<0,
  \qquad
  C_{Ht}<0,
  \qquad
  C_{Hq}^{(+)}\simeq 0.
  \label{eq:dir}
\end{equation}
The first two conditions are favored by the correlated top rates, while the last is fixed by the LEP and SLC measurements of $Z\bar b_Lb_L$ included in the fit. At tree level, these current operators can be generated by mixing the top with vectorlike quarks (VLQs) or by exchanging spin-one states coupled to the Higgs and fermion currents. Scalar contributions to these operators first arise at loop level.

The VLQ multiplets and spin-one resonances that contribute at tree level to the electroweak current coefficients of interest are summarized in table~\ref{tab:vlq}, following the dictionary of Ref.~\cite{deBlas:2017xtg}.

\begin{table}[!htbp]
  \centering
  \renewcommand{\arraystretch}{1.25}
  \setlength{\tabcolsep}{1.5em}
  \begin{tabular}{@{}cccc@{}}
    \toprule
    State
      & $C_{Hq}^{(-)}$ & $C_{Hq}^{(+)}$ & $C_{Ht}$ \\
    \midrule
    \multicolumn{4}{@{}l}{\itshape
      Vectorlike quarks, $\times\,|y_\Psi|^2/M_\Psi^2$} \\
    \addlinespace[2pt]
    $\mathcal U\sim\mathbf 1_{2/3}$
      & $+1/2$ & $0$ & $0$ \\
    $\mathcal D\sim\mathbf 1_{-1/3}$
      & $0$ & $-1/2$ & $0$ \\
    $\mathcal Q_1\sim\mathbf 2_{1/6}$
      & $0$ & $0$ & $-1/2$ \\
    $\mathcal Q_7\sim\mathbf 2_{7/6}$
      & $0$ & $0$ & $+1/2$ \\
    $\mathcal T_1\sim\mathbf 3_{-1/3}$
      & $-1/4$ & $-1/8$ & $0$ \\
    $\mathcal T_2\sim\mathbf 3_{2/3}$
      & $+1/8$ & $+1/4$ & $0$ \\
    \midrule
    \multicolumn{4}{@{}l}{\itshape
      Spin-one resonances,
      $\times\,g_{\mathcal V}^H g_{\mathcal V}^{q,t}/M_{\mathcal V}^2$} \\
    \addlinespace[2pt]
    $\mathcal W\sim\mathbf 3_0$   & $+1/4$ & $-1/4$ & $0$    \\
    $\mathcal B\sim\mathbf 1_0$   & $-1$   & $-1$   & $-1$   \\
    \bottomrule
  \end{tabular}
  \caption{Tree level contributions to the third-generation electroweak current coefficients, from the dictionary of Ref.~\cite{deBlas:2017xtg}, in terms of $C_{Hq}^{(\pm)}=C_{Hq}^{(1)}\pm C_{Hq}^{(3)}$. The upper block lists the color triplet VLQ multiplets $\Psi$, labeled by their $SU(2)_L$ representation and hypercharge, with $y_\Psi$ the VLQ Higgs coupling. The lower block lists the color singlet spin-one resonances, a $W'$ and a $Z'$, with $g_{\mathcal V}^H$ their coupling to the Higgs current and $g_{\mathcal V}^{q,t}$ to the quark currents. Each entry is the factor multiplying the combination quoted in the corresponding block heading. All entries assume the coupling normalization of that reference, in which the $\mathcal W$ interactions carry an explicit factor $1/2$ and the $\mathcal B$ ones do not, so the relative factor of four between the two spin-one rows follows from that choice rather than from the dynamics.}
  \label{tab:vlq}
\end{table}

No single state of table~\ref{tab:vlq} reproduces Eq.~\eqref{eq:dir}, since every state that can give a negative $C_{Hq}^{(-)}$ also shifts $C_{Hq}^{(+)}$. The preferred pattern therefore requires a combination of states.

Including these states without further structure breaks the custodial symmetry of the SM scalar sector, since their couplings to the Higgs do not respect it in general. This breaking is constrained by the oblique parameter $T$, with $\Delta\rho=\alpha(M_Z^2)T$ in the Peskin--Takeuchi formalism \cite{Peskin:1991sw}. Profiling over $S$ and setting $U=0$, the 2025 PDG update gives~\cite{ParticleDataGroup:2024cfk}
\begin{equation}
  T=0.047\pm0.066.
  \label{eq:Tfit}
\end{equation}
The narrow allowed range motivates imposing a custodial symmetry on the NP sector responsible for observable shifts in the top electroweak couplings. We take the custodial group to be $SO(4)\simeq SU(2)_L\times SU(2)_R$ and supplement it by $U(1)_X$ so that hypercharge is embedded as $Y=T_R^3+X$.

\subsection{An effective composite Higgs model}
\label{sec:model}
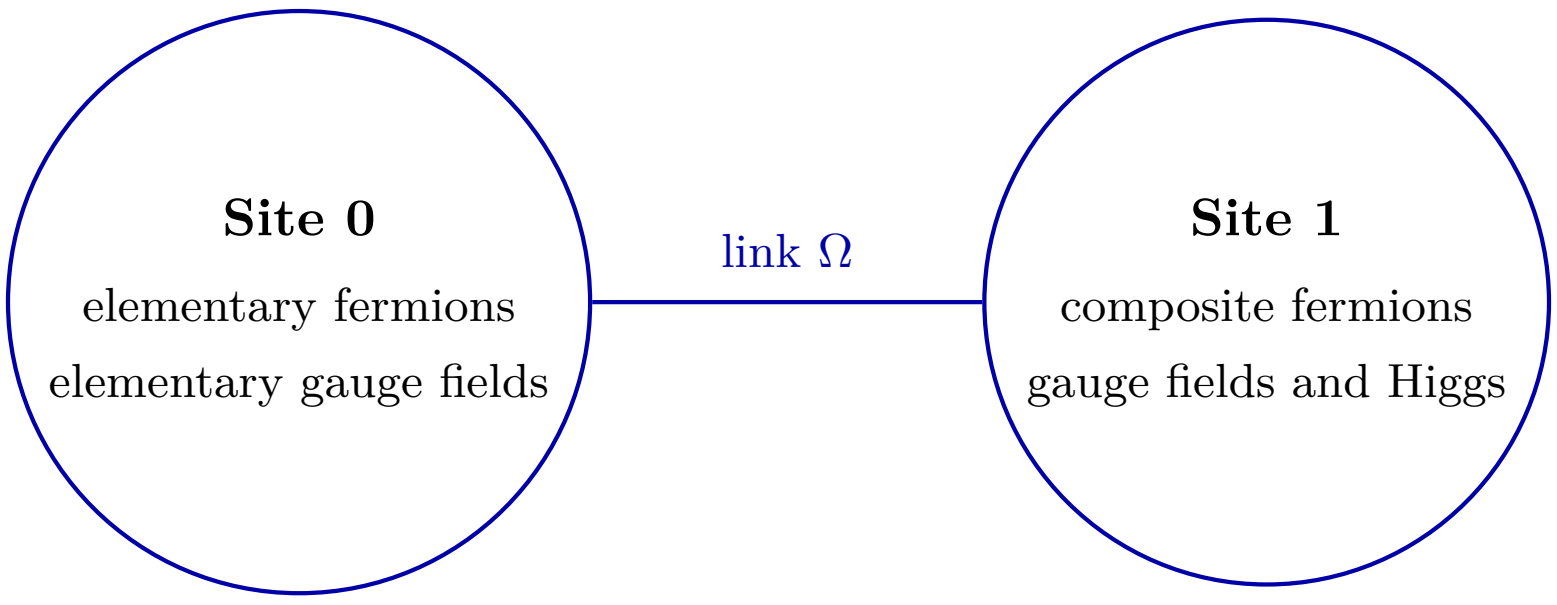
\begin{figure}[!htbp]
  \centering
  \begin{tikzpicture}[
    site/.style={circle, draw=blue!65!black, line width=0.8pt,
      minimum size=2.65cm, align=center, inner sep=3pt, font=\small},
    mixing/.style={blue!65!black, line width=0.8pt}
  ]
    \node[site] (el) at (0,0) {%
      {\bfseries Site 0}\\[0.3em]
      {\footnotesize elementary fermions}\\[0.1em]
      {\footnotesize elementary gauge fields}%
    };
    \node[site] (comp) at (6.4,0) {%
      {\bfseries Site 1}\\[0.3em]
      {\footnotesize composite fermions}\\[0.1em]
      {\footnotesize gauge fields and Higgs}%
    };
    \draw[mixing] (el.east) --
      node[above=2.5pt, fill=white, inner xsep=3pt,
        font=\footnotesize] {link $\Omega$}
      (comp.west);
  \end{tikzpicture}
  \caption{Picture of the two-site model, with the link field $\Omega$ connecting both sectors.}
  \label{fig:2site}
\end{figure}

New VLQs and spin-one resonances, as well as custodial symmetry, arise naturally in a composite Higgs framework, whose lowest lying states can be described by a two-site model~\cite{Contino:2006nn}, Fig.~\ref{fig:2site}. Site 0 contains the elementary fields, which match the gauge symmetries and fermion content of the SM, whereas site 1 contains the first level of resonances of the composite sector, including the Higgs field.\footnote{In this article the Higgs is taken as an ordinary scalar field at site 1, but it is straightforward to extend the symmetry of site 1 to make the Higgs a pseudo Nambu Goldstone boson (pNGB)~\cite{DeCurtis:2011yx,Carena:2014ria}.} The composite spin-one resonances are included by considering a gauge symmetry at site 1, under which the matter fields can be charged. A nonlinear sigma field $\Omega$, transforming under the gauge symmetries of both sites as $\Omega\to U_{\mathrm{el}}\Omega U_{\mathrm{cp}}^\dagger$, connects them. All the couplings at site 1 are taken to be large but perturbative, $1<g_*\ll4\pi$, and the scale of the resonances is taken to be of order a few TeV~\cite{Contino:2006nn,Contino:2006qr,DeCurtis:2011yx,Agashe:2008uz}.

To account for custodial symmetry, the symmetry at site 1 is $[SU(2)_L\times SU(2)_R\times U(1)_X]_1$.\footnote{There is also an $SU(3)_c$ factor at each site, whose resonance does not enter the electroweak current coefficients considered here and is not included in our analysis.} We denote its representations by $(\mathbf a,\mathbf b)_X$, with the subscript giving the $U(1)_X$ charge. The Higgs transforms as a bidoublet $\mathcal H\sim(\mathbf 2,\mathbf 2)_0$, built from the SM doublet $H$ and its conjugate $\widetilde H$.

The gauge fields at site 1 are $\rho_{L\mu}^a$, $\rho_{R\mu}^a$ and $\rho_{X\mu}$. Under $SU(2)_L\times U(1)_Y$ they decompose into the triplet $\rho_{L\mu}^a\sim\mathbf 3_0$, the neutral singlets $\rho_{R\mu}^3$ and $\rho_{X\mu}\sim\mathbf 1_0$, and the charged singlets $\rho_{R\mu}^\pm\sim\mathbf 1_{\pm1}$. These are the spin-one resonances of the two-site construction. Their couplings $g_L^*$, $g_R^*$ and $g_X^*$ are independent in general, and we adopt the simplifying choice $g_X^*=g_R^*$ of Ref.~\cite{Contino:2006nn}. Among the spin-one states of table~\ref{tab:vlq}, $\rho_L$ plays the role of the vector triplet and each neutral singlet that of the vector singlet, while the charged $\rho_R^\pm$ do not contribute to the coefficients considered here.

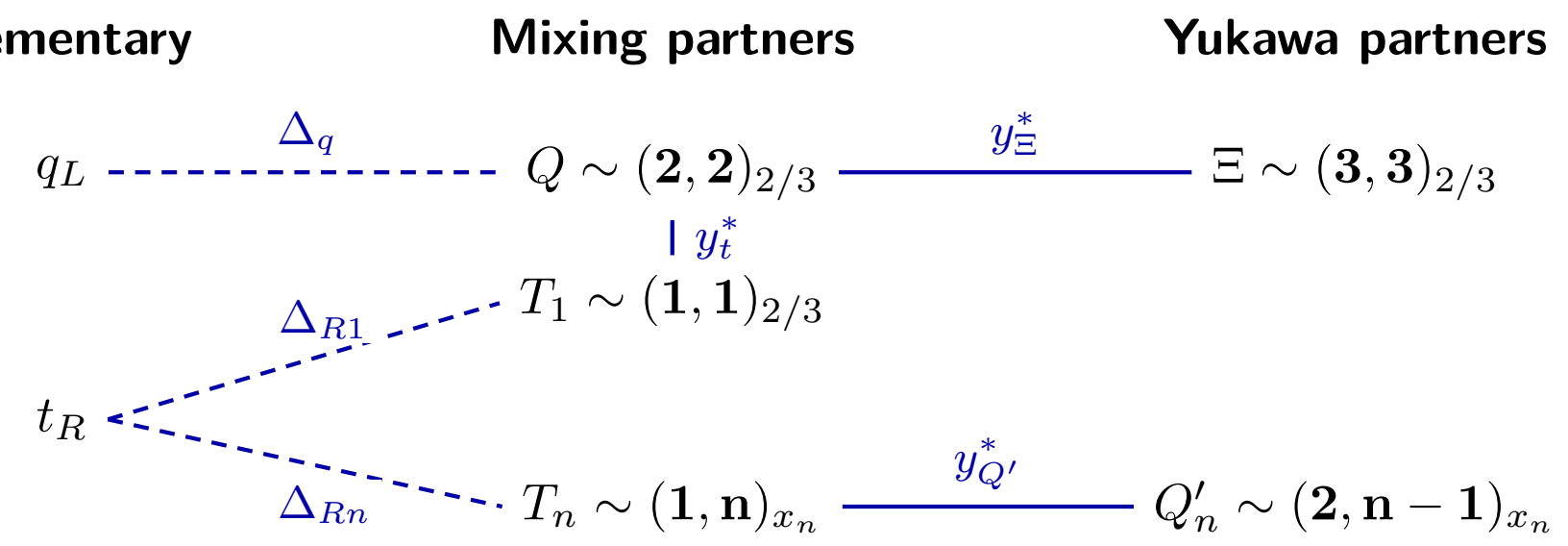
\begin{figure}[!htbp]
  \centering
  \begin{tikzpicture}[
    heading/.style={font=\small\sffamily\bfseries},
    field/.style={font=\small},
    mixing/.style={blue!65!black, dashed, dash pattern=on 3pt off 2.2pt,
      line width=0.8pt},
    yukawa/.style={blue!65!black, line width=0.8pt},
    label/.style={fill=white, inner xsep=2pt, inner ysep=1pt,
      font=\footnotesize}
  ]
    \node[heading] at (0.25,1.55) {Elementary};
    \node[heading] at (4.45,1.55) {Mixing partners};
    \node[heading] at (9.15,1.55) {Yukawa partners};

    \node[field] (q) at (0.25,0.65) {$q_L$};
    \node[field] (t) at (0.25,-1.05) {$t_R$};
    \node[field] (Q) at (4.45,0.65)
      {$Q\sim(\mathbf 2,\mathbf 2)_{2/3}$};
    \node[field] (T1) at (4.45,-0.25)
      {$T_1\sim(\mathbf 1,\mathbf 1)_{2/3}$};
    \node[field] (Tn) at (4.45,-1.65)
      {$T_n\sim(\mathbf 1,\mathbf n)_{x_n}$};
    \node[field] (X) at (9.15,0.65)
      {$\Xi\sim(\mathbf 3,\mathbf 3)_{2/3}$};
    \node[field] (Qn) at (9.15,-1.65)
      {$Q'_n\sim(\mathbf 2,\mathbf{n-1})_{x_n}$};

    \draw[mixing] (q.east) --
      node[label,above=2pt] {$\Delta_q$} (Q.west);
    \draw[mixing] (t.east) --
      node[label,above=2pt,pos=0.55] {$\Delta_{R1}$} (T1.west);
    \draw[mixing] (t.east) --
      node[label,below=2pt,pos=0.55] {$\Delta_{Rn}$} (Tn.west);
    \draw[yukawa] (Q.south) --
      node[label,right=2pt] {$y_t^*$} (T1.north);
    \draw[yukawa] (Q.east) --
      node[label,above=2pt] {$y_\Xi^*$} (X.west);
    \draw[yukawa] (Tn.east) --
      node[label,above=2pt] {$y_{Q'}^*$} (Qn.west);
  \end{tikzpicture}
  \caption{Fermion content of the two-site model. Dashed lines denote elementary/composite mixing, while solid lines denote Yukawa couplings within the composite sector.}
  \label{fig:ferm}
\end{figure}

Associated with each elementary quark there is a set of composite VLQs charged under $SU(2)_R$. We do not consider partners of the light fermions of the first and second generations, and $b_R$, but they can be included straightforwardly~\cite{Agashe:2004cp}. The composite fermion content at site 1, shown in Fig.~\ref{fig:ferm}, consists of a bidoublet $Q\sim(\mathbf 2,\mathbf 2)_{2/3}$, a bitriplet $\Xi\sim(\mathbf 3,\mathbf 3)_{2/3}$, two $SU(2)_L$ singlets $T_1\sim(\mathbf 1,\mathbf 1)_{2/3}$ and $T_n\sim(\mathbf 1,\mathbf n)_{x_n}$ with $n\geq2$, and an $SU(2)_L$ doublet $Q'_n\sim(\mathbf 2,\mathbf{n-1})_{x_n}$.

Such a model has the Lagrangian
\begin{equation}
  \mathcal L=\mathcal L_{\mathrm{el}}+\mathcal L_{\mathrm{cp}}
  +\mathcal L_{\mathrm{mix}},
\end{equation}
where the elementary sector contains the elementary gauge fields and third-generation quarks,
\begin{equation}
  \mathcal L_{\mathrm{el}}=
  -\frac14 F_{\mu\nu}^2
  +\bar q_L^{\mathrm{el}}i\!\not\!\!D q_L^{\mathrm{el}}
  +\bar t_R^{\mathrm{el}}i\!\not\!\!D t_R^{\mathrm{el}},
\end{equation}
the composite sector contains the resonances, the Higgs and their interactions,
\begin{align}
  \mathcal L_{\mathrm{cp}}={}&
  -\frac14 \rho_{\mu\nu}^2+|D_\mu^{\mathrm{cp}}\mathcal H|^2-V(\mathcal H)
  +\sum_\Psi\bar\Psi(i\!\not\!\!D^{\mathrm{cp}}-M_\Psi)\Psi
  \notag\\
  &-\Bigl[
    y_t^*\bar Q_L\mathcal H T_{1R}
    +y_\Xi^*\bar Q_L\mathcal H\Xi_R
    +y_{Q'}^*\bar Q'_{nL}\mathcal H T_{nR}
    \Bigr]+\text{h.c.},
  \label{eq:Lcp}
\end{align}
and the link field connects the two,
\begin{align}
  \mathcal L_{\mathrm{mix}}={}&
  \frac{f^2}{4}|D_\mu\Omega|^2
  +\Bigl[\Delta_q\bar q_L^{\mathrm{el}}\Omega Q_{1/6,R}
  +\Delta_{R1}\bar T_{1L}\Omega^\dagger t_R^{\mathrm{el}}
  +\Delta_{Rn}\overline{T}_{n,2/3,L} \Omega^\dagger t_R^{\mathrm{el}}
  \Bigr]+\text{h.c.}
  \label{eq:Lmix}
\end{align}
Here $F_{\mu\nu}$ and $\rho_{\mu\nu}$ are the elementary and composite field strengths, $D_\mu$ and $D_\mu^{\mathrm{cp}}$ the covariant derivatives of each site, and the $\Delta$'s the elementary/composite mass mixings. The link field is charged under the symmetries of both sites, so $D_\mu\Omega$ involves both sets of gauge fields, and $f$ is its decay constant. Each factor of the composite group can be connected by its own link field with an independent decay constant~\cite{DeCurtis:2011yx}, but for simplicity we take a common $f$. The sum runs over $\Psi=Q,\Xi,T_1,T_n,Q'_n$. Subscripts on a composite multiplet denote the hypercharge of the component, as in $Q_{1/6}$ and $T_{n,2/3}$. The Yukawa interactions shown above are custodially invariant. The corresponding wrong chirality interactions, obtained by interchanging $L$ and $R$, are also allowed by the symmetry. They do not contribute to the dimension-six tree level matching of the electroweak current operators considered here, but they modify the fermionic one-loop contributions to $S$ and $T$. To keep the fermion sector minimal, we set these couplings to zero.\footnote{In pNGB completions of the two-site model, such interactions also spoil the one-loop finiteness of the Higgs potential, introducing a logarithmic divergence~\cite{DeCurtis:2011yx}.}

In unitary gauge, $\Omega=1$. The link kinetic term generates the elementary/composite gauge mixing, leaving a massless diagonal combination identified with the SM fields and giving mass to the orthogonal spin-one state, while the fermionic terms containing $\Omega$ reduce to ordinary mass mixings. We write $\langle H\rangle=v_H/\sqrt2$. Since the Higgs couples to the elementary $W$ only through the link fields, the Fermi constant obtained by integrating out the composite vectors at tree level gives $v^{-2}=v_H^{-2}+f^{-2}$. Hence $v_H$ is slightly larger than $v\simeq246~\mathrm{GeV}$ and $v/v_H=\sqrt{1-v^2/f^2}$.

We discuss now the role and properties of each VLQ. Protecting $Z\bar b_L b_L$ requires the composite state that mixes with $b_L$ to be an eigenstate of the custodial parity $P_{LR}$, which exchanges $SU(2)_L$ and $SU(2)_R$. This requires $T_L=T_R$ and $T_L^3=T_R^3$~\cite{Agashe:2006at}, fixing the embedding of $q_L$ into the $Y=1/6$ doublet of $Q\sim(\mathbf 2,\mathbf 2)_{2/3}$. Up to corrections suppressed by gauge mixing, this protection implies $C_{Hq}^{(+)}\simeq0$. Since $Q$ is a doublet of the composite $SU(2)_L$, both its current and the Higgs current couple to $\rho_L$ through the same composite gauge interaction, making the product of couplings positive. Together with the $+1/4$ entry of the vector triplet row in table~\ref{tab:vlq}, this gives a positive contribution to $C_{Hq}^{(-)}$, opposite to the direction required by Eq.~\eqref{eq:dir}. The vector singlet contributes with the same sign to $C_{Hq}^{(-)}$ and cancels the triplet contribution to $C_{Hq}^{(+)}$, as required by $P_{LR}$.

A negative fermionic contribution to $C_{Hq}^{(-)}$ is therefore needed. In the upper block of table~\ref{tab:vlq}, the only multiplet with a negative $C_{Hq}^{(-)}$ entry is the weak triplet $\mathcal T_1$, which also shifts $C_{Hq}^{(+)}$. The custodially complete bitriplet $\Xi$ contains both weak triplets of the table: $\mathcal T_1$ and $\mathcal T_2$, with their relative couplings fixed by the symmetry. Their combined contribution provides the required negative shift in $C_{Hq}^{(-)}$ while preserving the protection of $Z\bar b_L b_L$.

In the right-handed sector, the top Yukawa requires an invariant coupling between $Q$, the Higgs and the composite partner of $t_R$. Since $t_R$ is an $SU(2)_L$ singlet, the group structure leaves only $(\mathbf 1,\mathbf 1)_{2/3}$ and $(\mathbf 1,\mathbf 3)_{2/3}$. We choose the former, $T_1$, for simplicity.\footnote{The $P_{LR}$ symmetric alternative is $(\mathbf 1,\mathbf 3)_{2/3}\oplus(\mathbf 3,\mathbf 1)_{2/3}$. It requires a larger multiplet and, for a fixed top Yukawa, enhances the positive fermionic contribution to $C_{Hq}^{(-)}$, opposite to the direction preferred by the fit.} In either case, the component that mixes with $t_R$ has $T_L^3=T_R^3=0$ and is therefore protected by the custodial symmetry $P_C$~\cite{Agashe:2006at}. The leading correction to $Z\bar t_Rt_R$, and correspondingly to $C_{Ht}$, vanishes in the custodial limit. Custodial breaking from the elementary gauging and mixings leaves a suppressed positive contribution.

Obtaining $C_{Ht}<0$ therefore requires a second embedding of $t_R$ that is not protected by $P_C$. This is the role of $T_n$. Placing its charge $2/3$ component at the highest weight, $T_R^3=(n-1)/2$, fixes
\begin{equation}
  x_n=\frac{2}{3}-\frac{n-1}{2}.
\end{equation}
The coupling product entering the vector singlet row of table~\ref{tab:vlq} is proportional to $(n-1)/2$. Since the Higgs is $X$ neutral, only $\rho_R^3$ is exchanged.

The $-1$ entry of the vector singlet row then gives a negative contribution to $C_{Ht}$, whose magnitude grows with $n$ at fixed mixing.

The Yukawa coupling of $T_n$ through the Higgs allows partners transforming as $(\mathbf 2,\mathbf{n-1})_{x_n}$ or $(\mathbf 2,\mathbf{n+1})_{x_n}$. We choose the smaller representation, $Q'_n\sim(\mathbf 2,\mathbf{n-1})_{x_n}$, whose largest hypercharge is $1/6$ for every $n\geq2$. With this choice $Q'_n$ contains a state $\mathcal Q_1$, with the $-1/2$ entry in table~\ref{tab:vlq}, while avoiding the oppositely signed state $\mathcal Q_7$ contained in the larger representation. Hence the spin-one and fermion exchanges both shift $C_{Ht}$ toward negative values.

\subsection{Bases and rotations}
\label{sec:rot}

We match onto the SMEFT in the electroweak symmetric basis, in which the elementary/composite mixings are diagonalized. The additional rotation after electroweak symmetry breaking gives the physical mass basis. It is not needed for the tree level matching but is required to determine the spectrum and observables such as the oblique parameters. Both rotations are given in Appendices~\ref{app:ferm} and \ref{app:vec}.

With $g_X^*=g_R^*$, the neutral composite fields separate into the direction $\rho^\parallel=(\rho_R^3+\rho_X)/\sqrt2$ parallel to hypercharge and the orthogonal state $\rho^\perp=(\rho_R^3-\rho_X)/\sqrt2$. The link mixing is then removed, leaving $W_\mu^a$ and $B_\mu$ as the light SM states and $W_\mu^{*a}$ and $B_\mu^*$ as heavy. The fields $\rho_\mu^\perp$ and $\rho_{R\mu}^\pm$ have no elementary partners and are already heavy eigenstates. Writing $g$ and $g'$ for the low energy $SU(2)_L$ and $U(1)_Y$ gauge couplings, the sines and cosines of the mixing angles are
\begin{equation}
  s_L=\frac{g}{g_L^*},
  \qquad
  s_Y=\frac{\sqrt2\,g'}{g_R^*},
  \qquad
  c_{L,Y}=\sqrt{1-s_{L,Y}^2},
\end{equation}
and the heavy vector masses are
\begin{equation}
  M_{W^*}^2=\frac{f^2g_L^{*2}}{2c_L^2},
  \qquad
  M_{B^*}^2=\frac{f^2g_R^{*2}}{2c_Y^2},
  \qquad
  M_{\rho^\perp}^2=M_{\rho_R^\pm}^2=\frac{f^2g_R^{*2}}{2}.
  \label{eq:mV}
\end{equation}
The sines $s_L$ and $s_Y$ measure the compositeness of the SM $W$ and $B$, and are small since $g_L^*$ and $g_R^*$ are much larger than $g$ and $g'$.

In the fermion sector the massless states are
\begin{equation}
  \begin{gathered}
    q_L^{(0)}=c_q q_L^{\mathrm{el}}+s_qQ_{1/6,L},
    \qquad
    t_R^{(0)}=c_\chi t_R^{\mathrm{el}}+a_1T_{1R}+a_nT_{nR},\\
    a_1=s_\chi c_\phi,\qquad a_n=s_\chi s_\phi,\qquad a_1^2+a_n^2=s_\chi^2.
  \end{gathered}
  \label{eq:light}
\end{equation}
Each angle is fixed by the ratio of the corresponding mixing to the composite mass, $\tan\theta_q=\Delta_q/M_Q$ for the doublet and $\tan\chi\cos\phi=\Delta_{R1}/M_{T_1}$, $\tan\chi\sin\phi=\Delta_{Rn}/M_{T_n}$ for the singlets. As for the vectors, $s_q$ and $s_\chi$ measure the compositeness of $q_L$ and $t_R$, with $\phi$ distributing the composite component between the two embeddings through $a_1$ and $a_n$.

The bitriplet $\Xi$ and the doublet $Q'_n$ have no elementary partners and are already eigenstates of this basis. The $Q$ masses are $m_{Q_{1/6}}=M_Q/c_q$ and $m_{Q_{7/6}}=M_Q$, and the two $t_R$ mixings leave the heavy $T_1$, $T_n$ block non-diagonal. It need not be explicitly diagonalized for the matching, since the current operators involve a single heavy propagator and are insensitive to rotations within that block.

The top mass is
\begin{equation}
  m_t=\frac{y_t^*v_H}{\sqrt{2}}s_qa_1+\mathcal O(v_H^3/M^2).
\end{equation}
Reproducing $m_t$ with a perturbative $y_t^*$ fixes $s_qa_1$ to be of order $1/y_t^*$, so at least one of the two mixings is sizable and neither can be arbitrarily small.

\subsection{Matching the top electroweak couplings}
Applying the rotations of Sec.~\ref{sec:rot} and integrating out the heavy states at tree level, the electroweak current coefficients receive contributions mediated by spin-one exchange and from Yukawa interactions, hereafter referred to as $V$ and $F$, respectively,
\begin{equation}
  C_i=C_i^V+C_i^F,
  \qquad i=Hq^{(-)},\ Hq^{(+)},\ Ht.
\end{equation}
Projecting the light fermions onto the currents collected in Appendix~\ref{app:vec} and applying the tree level dictionary of Ref.~\cite{deBlas:2017xtg} gives
  \begin{align}
  C_{Hq}^{(-),V}
  &=\frac{(1+c_L^2)s_q^2-c_L^2s_L^2}{2f^2}
    +\frac{s_Y^2(c_Y^2+s_q^2)}{12f^2},
  \label{eq:CV3}\\
  C_{Hq}^{(+),V}
  &=\frac{1}{12f^2}\left[
  s_Y^2\left(c_Y^2+s_q^2\right)
  +6s_L^2\left(c_L^2+s_q^2\right)\right],
  \label{eq:CVp}\\
  C_{Ht}^{V}
  &=\frac{s_Y^2}{3f^2}
    \left(c_Y^2+s_\chi^2\right)
    -\frac{(n-1)a_n^2}{2f^2}.
  \label{eq:CVR}
\end{align}
As discussed in Sec.~\ref{sec:model}, the vector singlet terms follow $-T_R^3$ of the state the quark mixes with, and $T_R^3$ is $-1/2$ for $Q_{1/6}$ and $(n-1)/2$ for $T_n$. This gives the singlet contributions to $C_{Hq}^{(\pm),V}$ and the $T_n$ term of $C_{Ht}^{V}$ opposite signs. The latter is negative, grows with $n$ and is unsuppressed by the elementary/composite gauge mixing.

The fermionic contributions follow from the same dictionary, with the normalization of Eq.~\eqref{eq:Lcp},
\begin{align}
  C_{Hq}^{(-),F}
  &=\frac{|y_t^*|^2s_q^2}{2}
    \left[\frac{(1-a_1^2)^2}{M_{T_1}^2}
    +\frac{a_1^2a_n^2}{M_{T_n}^2}\right]
    -\frac{3|y_\Xi^*|^2s_q^2}{8M_\Xi^2},
  \qquad C_{Hq}^{(+),F}=0,
  \label{eq:CFL}\\
  C_{Ht}^{F}
  &=\frac{|y_t^*|^2a_1^2s_q^2}{2M_Q^2}(1+c_q^2)
    -\frac{|y_{Q'}^*|^2a_n^2}{2M_{Q'}^2}.
  \label{eq:CFR}
\end{align}
The bitriplet contribution can overcome the positive $y_t^*$ term in $C_{Hq}^{(-),F}$ without contributing to $C_{Hq}^{(+),F}$, and the $Q'_n$ channel shifts $C_{Ht}^{F}$ toward negative values.

Summing the two contributions, the coefficients take a simple form at leading order in the elementary/composite gauge mixing. Since $s_L=g/g_L^*$ and $s_Y=\sqrt2g'/g_R^*$, both mixings are small in the strongly coupled regime $g_{L,R}^*\gg g,g'$, and setting $s_{L,Y}\to0$ in Eqs.~\eqref{eq:CV3}--\eqref{eq:CFR} gives
\begin{align}
  C_{Hq}^{(-)}
  &\simeq s_q^2\left[\frac{1}{f^2}
    +\frac{|y_t^*|^2}{2}\left(\frac{(1-a_1^2)^2}{M_{T_1}^2}
    +\frac{a_1^2a_n^2}{M_{T_n}^2}\right)
    -\frac{3|y_\Xi^*|^2}{8M_\Xi^2}\right],
  \qquad
  C_{Hq}^{(+)}\simeq0,
  \notag\\
  C_{Ht}
  &\simeq-a_n^2\left[\frac{n-1}{2f^2}
    +\frac{|y_{Q'}^*|^2}{2M_{Q'}^2}\right]
    +\frac{|y_t^*|^2a_1^2s_q^2(1+c_q^2)}{2M_Q^2}.
  \label{eq:LO}
\end{align}
The spin-one contribution to $C_{Hq}^{(-)}$ is positive, so a negative $C_{Hq}^{(-)}$ requires the bitriplet term to exceed the other two, which requires a large $y_\Xi^*$, a light $\Xi$ and a moderate $y_t^*$. In $C_{Ht}$ the two negative contributions are both proportional to $a_n^2$, the fraction of $t_R$ in $T_n$, while the positive $y_t^*$ term is controlled by $a_1$.

\subsection{Custodial protection and electroweak precision observables}
\label{sec:ewpt}
The custodial structure of the composite sector protects the $T$ parameter, as well as the $Z\bar b_Lb_L$ coupling through $P_{LR}$.

The vanishing of $C_{Hq}^{(+),F}$ in Eq.~\eqref{eq:CFL} is the simplest manifestation of the $P_{LR}$ protection. More generally, $P_{LR}$ cancels every contribution to $C_{Hq}^{(+)}$ not suppressed by the gauge mixing. The elementary gauging of $SU(2)_L\times U(1)_Y$ breaks $P_{LR}$ and leaves only the vector contribution of Eq.~\eqref{eq:CVp}, so the residual shift of the $Z\bar b_Lb_L$ vertex is suppressed by $s_L^2$ and $s_Y^2$.

The gauge rotations of Sec.~\ref{sec:rot} also induce couplings of the heavy vectors to the elementary SM fermions. Their exchange generates universal contributions to $\mathcal O_{H\ell}^{(1)}$, $\mathcal O_{H\ell}^{(3)}$, $\mathcal O_{He}$, $\mathcal O_{Hq}^{(1)}$, $\mathcal O_{Hq}^{(3)}$, $\mathcal O_{Hu}$ and $\mathcal O_{Hd}$, together with correlated four-fermion interactions, including $\mathcal O_{\ell\ell}$. The universal corrections can be expressed through electroweak vacuum polarizations, whose dominant effects in the strongly coupled regime considered here are parametrized by $S$ and $T$ \cite{Contino:2006nn}. We compute the corresponding shifts $\Delta S$ and $\Delta T$ directly in the model and compare them with the electroweak precision fit.

At tree level the spin-one states generate $S_V^{\mathrm{tree}}$ of Eq.~\eqref{eq:Sv}, which decreases as $f$ and $g_{L,R}^*$ grow, and give $T_V^{\mathrm{tree}}=0$. The 2025 PDG fit~\cite{ParticleDataGroup:2024cfk} gives $S=0.026\pm0.075$, with a correlation of $0.90$ with the $T$ of Eq.~\eqref{eq:Tfit}, and bounds the masses of the electroweak spin-one resonances. The fermion sector generates oblique corrections at one loop~\cite{Albergaria:2023nby}, with explicit formulas given in Appendix~\ref{app:fST}. Retaining the lowest nonzero order in each resonance sector, we use
\begin{equation}
  \Delta S= S_V^{\mathrm{tree}}+\Delta S_F^{1\ell},
  \qquad
  \Delta T=\Delta T_F^{1\ell}.
  \label{eq:ST}
\end{equation}

\subsection{Further effects of the resonances}
\label{sec:other}
The effects of the new resonances are not limited to the top electroweak currents. Their mixing also modifies Higgs couplings, while spin-one exchange generates four-quark interactions.

We describe the modified Higgs couplings in the $\kappa$ framework~\cite{LHCHiggsCrossSectionWorkingGroup:2012nn}. Rotating the Yukawa and vector interactions to the physical basis gives
\begin{gather}
  \kappa_t=\frac{v}{m_t}
    \left[U_L^{2/3\dagger}Y_h^{2/3}U_R^{2/3}\right]_{tt},
  \qquad
  \kappa_b=\frac{v}{m_b}
    \left[U_L^{-1/3\dagger}Y_h^{-1/3}U_R^{-1/3}\right]_{bb},\\
  \kappa_W=\frac{v}{2m_W^2}
    \left[R_C^T G_h^C R_C\right]_{WW},
  \qquad
  \kappa_Z=\frac{v}{2m_Z^2}
    \left[R_N^T G_h^N R_N\right]_{ZZ}.
  \label{eq:kV}
\end{gather}
The Higgs coupling matrices are $Y_h^{\mathcal Q}=\partial\mathcal M_{\mathcal Q}^{(0)}/\partial v_H$ and $G_h^{C,N}=\partial\mathcal M_{C,N}^2/\partial v_H$, with the rotations given in Appendices~\ref{app:fphys} and \ref{app:vphys}.

Integrating out the composite fields gives form factors that encode the effective kinetic and mass terms of the light fields. Their zero momentum values provide simple estimates of the masses and Higgs couplings~\cite{Carena:2014ria}, with $\kappa_i^{(0)}=v\,\partial_{v_H}\log m_i^{(0)}$. Using a small elementary Yukawa to reproduce the bottom mass gives
\begin{align}
  \kappa_t^{(0)} &\simeq \frac{v}{v_H}
    \left( 1 - \frac{\epsilon_L}{1+\epsilon_L} - \frac{\epsilon_R}{1+\epsilon_R} \right),
  &\kappa_b^{(0)} &\simeq \frac{v/v_H}{1+\epsilon_b}\,,
  \label{eq:kF} \\[0.8ex]
  \kappa_W^{(0)} &\simeq \frac{v/v_H}{1+\epsilon_W}\,,
  &\kappa_Z^{(0)} &\simeq \frac{v/v_H}{1+\epsilon_Z}\,,
  \label{eq:kVapp}
\end{align}
where
\begin{gather*}
  \epsilon_L = s_q^2 v_H^2 \left(
    \frac{|y_t^*|^2}{2M_{T_1}^2} + \frac{5|y_\Xi^*|^2}{8M_\Xi^2} \right), \qquad
  \epsilon_R = v_H^2 \left(
    \frac{a_1^2 |y_t^*|^2}{M_Q^2} + \frac{a_n^2 |y_{Q'}^*|^2}{2M_{Q'}^2} \right), \\[1ex]
  \epsilon_b = \frac{s_q^2 v_H^2 |y_\Xi^*|^2}{2M_\Xi^2}\,, \\[1ex]
  \epsilon_W = c_L^2 \frac{v_H^2}{f^2}\,, \qquad
  \epsilon_Z = \left( c_L^2 - \frac{s_Y^2}{2} \right) \frac{v_H^2}{f^2}\,.
\end{gather*}

The numerical results use the full rotations rather than these approximations.

The Higgs couplings to gluons and photons are parametrized by $\kappa_g$ and $\kappa_\gamma$. In our model, the top and its partners, and likewise the $W$ and the charged spin-one resonances in the photon amplitude, give the common rescaling $v/v_H$ in the heavy mass limit \cite{Carena:2014ria,Gillioz:2012se}. We evaluate both couplings with the full loop functions.

In the charge $-1/3$ sector, mixing through $y_\Xi^*$ reduces $\kappa_b$ and adds a heavy fermion contribution to the gluon amplitude. Since the light bottom loop is small, this can raise $\kappa_g$ above one, while the smaller electric charge suppresses the effect on $\kappa_\gamma$. A composite embedding of $b_R$, not included in this work, can generate additional contributions to these couplings and to $\kappa_b$.

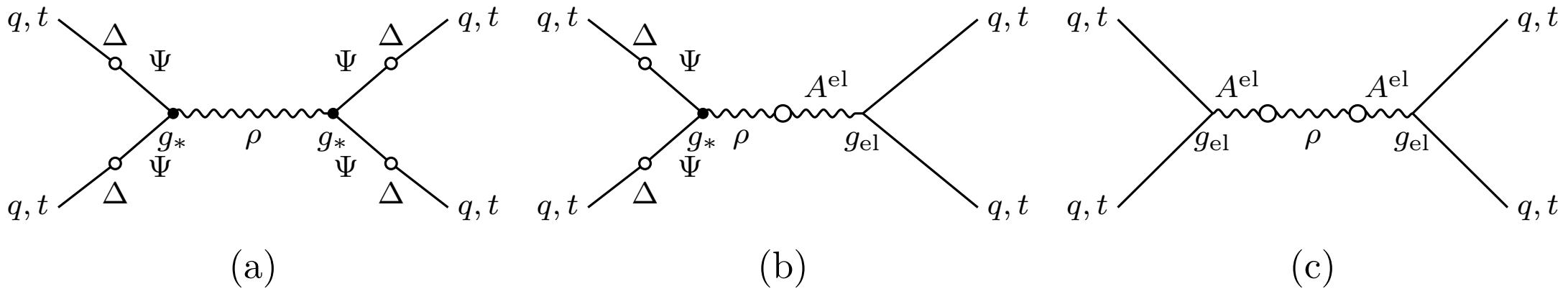
\begin{figure}[!htbp]
  \centering
  \begin{tikzpicture}[
      line width=0.6pt, scale=0.92,
      vec/.style={decorate,decoration={snake,amplitude=1pt,segment length=4.6pt}},
      vtx/.style={circle,fill,inner sep=1.05pt},
      mix/.style={circle,draw,fill=white,inner sep=1.05pt},
      lbl/.style={font=\scriptsize,inner sep=1.4pt},
      pan/.style={font=\small}
    ]
    \begin{scope}[xshift=0cm]
    \coordinate (L) at (-0.80,0); \coordinate (R) at (0.80,0);
    \draw (L) -- (-1.38, 0.50) node[mix]{} -- (-1.95, 0.94);
    \draw (L) -- (-1.38,-0.50) node[mix]{} -- (-1.95,-0.94);
    \draw[vec] (L) -- (R);
    \draw (R) -- ( 1.38, 0.50) node[mix]{} -- ( 1.95, 0.94);
    \draw (R) -- ( 1.38,-0.50) node[mix]{} -- ( 1.95,-0.94);
    \node[vtx] at (L) {}; \node[vtx] at (R) {};
    \node[lbl,anchor=east]  at (-1.98, 0.94) {$q,t$};
    \node[lbl,anchor=east]  at (-1.98,-0.94) {$q,t$};
    \node[lbl,anchor=west]  at ( 1.98, 0.94) {$q,t$};
    \node[lbl,anchor=west]  at ( 1.98,-0.94) {$q,t$};
    \node[lbl,anchor=north] at ( 0.00,-0.10) {$\rho$};
    \node[lbl,anchor=south west] at (-1.12, 0.36) {$\Psi$};
    \node[lbl,anchor=north west] at (-1.12,-0.36) {$\Psi$};
    \node[lbl,anchor=south east] at ( 1.12, 0.36) {$\Psi$};
    \node[lbl,anchor=north east] at ( 1.12,-0.36) {$\Psi$};
    \node[lbl,anchor=north] at (-0.80,-0.12) {$g_*$};
    \node[lbl,anchor=north] at ( 0.80,-0.12) {$g_*$};
    \node[lbl,anchor=south] at (-1.38, 0.62) {$\Delta$};
    \node[lbl,anchor=north] at (-1.38,-0.62) {$\Delta$};
    \node[lbl,anchor=south] at ( 1.38, 0.62) {$\Delta$};
    \node[lbl,anchor=north] at ( 1.38,-0.62) {$\Delta$};
    \node[pan] at (0,-1.55) {(a)};
    \end{scope}
    \begin{scope}[xshift=5.3cm]
      \coordinate (L) at (-0.80,0); \coordinate (R) at (0.80,0);
      \draw (L) -- (-1.38, 0.50) node[mix]{} -- (-1.95, 0.94);
      \draw (L) -- (-1.38,-0.50) node[mix]{} -- (-1.95,-0.94);
      \draw[vec] (L) -- (0,0); \draw[vec] (0,0) -- (R);
      \draw (R) -- (1.95, 0.94);
      \draw (R) -- (1.95,-0.94);
      \node[vtx] at (L) {}; \node[mix,inner sep=1.5pt] at (0,0) {};
      \node[lbl,anchor=east] at (-1.98, 0.94) {$q,t$};
      \node[lbl,anchor=east] at (-1.98,-0.94) {$q,t$};
      \node[lbl,anchor=west] at (1.98, 0.94) {$q,t$};
      \node[lbl,anchor=west] at (1.98,-0.94) {$q,t$};
      \node[lbl,anchor=south west] at (-1.12, 0.36) {$\Psi$};
      \node[lbl,anchor=north west] at (-1.12,-0.36) {$\Psi$};
      \node[lbl,anchor=south] at (-1.38, 0.62) {$\Delta$};
      \node[lbl,anchor=north] at (-1.38,-0.62) {$\Delta$};
      \node[lbl,anchor=north] at (-0.42,-0.10) {$\rho$};
      \node[lbl,anchor=south] at (0.44,0.10) {$A^{\mathrm{el}}$};
      \node[lbl,anchor=north] at (-0.80,-0.12) {$g_*$};
      \node[lbl,anchor=north] at (0.80,-0.12) {$g_{\mathrm{el}}$};
      \node[pan] at (0,-1.55) {(b)};
    \end{scope}
    \begin{scope}[xshift=10.6cm]
      \coordinate (L) at (-1.00,0); \coordinate (R) at (1.00,0);
      \draw (L) -- (-1.95, 0.94);
      \draw (L) -- (-1.95,-0.94);
      \draw (R) -- (1.95, 0.94);
      \draw (R) -- (1.95,-0.94);
      \draw[vec] (L) -- (-0.45,0);
      \draw[vec] (-0.45,0) -- (0.45,0);
      \draw[vec] (0.45,0) -- (R);
      \node[mix,inner sep=1.5pt] at (-0.45,0) {};
      \node[mix,inner sep=1.5pt] at (0.45,0) {};
      \node[lbl,anchor=east] at (-1.98, 0.94) {$q,t$};
      \node[lbl,anchor=east] at (-1.98,-0.94) {$q,t$};
      \node[lbl,anchor=west] at (1.98, 0.94) {$q,t$};
      \node[lbl,anchor=west] at (1.98,-0.94) {$q,t$};
      \node[lbl,anchor=south] at (-0.75,0.10) {$A^{\mathrm{el}}$};
      \node[lbl,anchor=north] at (0,-0.10) {$\rho$};
      \node[lbl,anchor=south] at (0.75,0.10) {$A^{\mathrm{el}}$};
      \node[lbl,anchor=north] at (-1.00,-0.12) {$g_{\mathrm{el}}$};
      \node[lbl,anchor=north] at (1.00,-0.12) {$g_{\mathrm{el}}$};
      \node[pan] at (0,-1.55) {(c)};
    \end{scope}
  \end{tikzpicture}
  \caption{Contributions to the four-quark operators of Eq.~\eqref{eq:4f} in an expansion in elementary/composite mixing. The exchanged resonance couples to (a) two composite currents, (b) one composite and one elementary current, or (c) two elementary currents. Panel (b) also contributes with the currents interchanged. Filled dots denote composite interactions and open circles fermion or gauge mixing. All three contributions are included in the coefficients above. The state $\rho^\perp$, which has no elementary partner, contributes only through (a).}
  \label{fig:4q}
\end{figure}

The spin-one resonances that modify the top electroweak currents also mediate interactions between two quark currents. Their exchange, illustrated in Fig.~\ref{fig:4q}, generates
\begin{align}
  \mathcal O_{qq}^{(1)}
    &=(\bar q\gamma_\mu q)(\bar q\gamma^\mu q),
  &
  \mathcal O_{qt}^{(1)}
    &=(\bar q\gamma_\mu q)(\bar t\gamma^\mu t),
  \notag\\
  \mathcal O_{qq}^{(3)}
    &=(\bar q\gamma_\mu\tau^a q)(\bar q\gamma^\mu\tau^a q),
  &
  \mathcal O_{tt}
    &=(\bar t\gamma_\mu t)(\bar t\gamma^\mu t),
  \label{eq:4f}
\end{align}
with coefficients
\begin{align}
  C_{qq}^{(3)}
  &=-\frac{\left(s_q^2-s_L^2\right)^2}{4f^2},
  \label{eq:Cqq3}\\
  C_{qq}^{(1)}
  &=-\frac{1}{72f^2}\left[\left(s_q^2-s_Y^2\right)^2+49\,s_q^4\right],\\
  C_{tt}
  &=-\frac{2\left(s_\chi^2-s_Y^2\right)^2}{9f^2}-\frac{\beta_n^2}{2f^2},
  \label{eq:Ctt}\\
  C_{qt}^{(1)}
  &=-\frac{\left(s_q^2-s_Y^2\right)\left(s_\chi^2-s_Y^2\right)}{9f^2}
    +\frac{7\,s_q^2\beta_n}{6f^2},
\end{align}
where $\beta_n=-\tfrac23a_1^2+\left(n-\tfrac53\right)a_n^2$ parametrizes the coupling of $\rho^\perp$ to the right-handed top current. Thus the choice of an embedding that enhances the spin-one contribution to $C_{Ht}$ also affects the four-quark interactions. At fixed mixings, this dependence is quadratic in $n$ for $C_{tt}$ and linear for $C_{qt}^{(1)}$. Only color singlet structures appear, since the resonance associated with the $SU(3)_c$ factors of Sec.~\ref{sec:model} is not included.

\FloatBarrier

\section{Numerical results}
\label{sec:num}
We now test whether the model can realize the direction selected by the fit while remaining compatible with electroweak precision data and direct constraints. The relative fermion and spin-one contributions allow shifts in several directions in the $(C_{Hq}^{(-)},C_{Ht})$ plane. We orient the scan toward $C_{Hq}^{(-)}<0$ and $C_{Ht}<0$, the direction that motivated the field content. We consider $n=2,\dots,5$, that is $T_n=(\mathbf 1,\mathbf 2)_{1/6}$, $(\mathbf 1,\mathbf 3)_{-1/3}$, $(\mathbf 1,\mathbf 4)_{-5/6}$ and $(\mathbf 1,\mathbf 5)_{-4/3}$, and sample the parameters uniformly over the ranges
\begin{equation}
  \begin{gathered}
    s_q\in(0.60,0.92),
    \quad
    s_\phi\in(0.68,0.92),
    \quad
    y_t^*\in(1.5,3.5),
    \quad
    y_\Xi^*\in(4.3,2\pi),
    \quad
    y_{Q'}^*\in(2.5,6),\\
    M_Q,\,M_{T_1}\in(2.5,5)~\mathrm{TeV},
    \quad
    M_{T_n}\in(1.7,5)~\mathrm{TeV},
    \\
    M_\Xi\in(1.9,3.3)~\mathrm{TeV},
    \quad
    M_{Q'}\in(1.5,4)~\mathrm{TeV},\\
    f\in(0.8,1.25)~\mathrm{TeV},
    \qquad
    g_L^*,\,g_R^*\in(4,2\pi).
  \end{gathered}
\end{equation}
We choose the lower endpoint $f=0.8~\mathrm{TeV}$ to keep the compositeness scale low, as favored by naturalness, while admitting points compatible with electroweak precision and Higgs coupling constraints. The Higgs vev $v_H$ is determined from $G_F$ through the relation in Sec.~\ref{sec:model}, and a small elementary Yukawa reproduces the bottom mass without introducing partners of $b_R$. At every point $s_\chi$ is fixed to reproduce the top mass, while $s_\phi$ sets how the compositeness of $t_R$ is shared between $T_1$ and $T_n$ through Eq.~\eqref{eq:light}. Sizable $s_q$ and $s_\phi$ favor appreciable shifts. Since the $y_t^*$ terms in Eq.~\eqref{eq:LO} shift both coefficients away from the target direction, we keep $y_t^*$ moderate and $M_Q,M_{T_1}$ comparatively large. Conversely, large $y_\Xi^*$ and $y_{Q'}^*$ enhance the desired shifts in $C_{Hq}^{(-)}$ and $C_{Ht}$. Pair production limits depend on the charge and decay pattern and typically lie between $1.3$ and $1.7~\mathrm{TeV}$ \cite{CMS:2024bni,ATLAS:2026ojf}. This motivates the lower ranges for $M_{T_n}$ and $M_{Q'}$. We use $1.5~\mathrm{TeV}$ as a common reference floor on the physical VLQ masses after diagonalization, supplemented by stronger bounds for the applicable decay channels.

In the region considered, points generating sizable negative shifts in $C_{Hq}^{(-)}$ with $M_\Xi<1.9~\mathrm{TeV}$ do not survive the CMS and ATLAS searches for single production of a charge $-4/3$ quark decaying into $Wb$, applied within the mass and width ranges they cover \cite{CMS:2026vwc,ATLAS:2025bzt}. The bitriplet contains such a state, and this motivates the lower boundary chosen for $M_\Xi$.

Large but perturbative $g_L^*$ and $g_R^*$ suppress the spin-one contribution to $S$. The upper coupling boundary is motivated by naive dimensional analysis \cite{Manohar:1983md}, with a nominal loop factor $g_*^2/(16\pi^2)\leq1/4$ before group and multiplicity factors. Corrections can therefore be appreciable, particularly for larger representations.

\subsection{Reach of the model}
We first ask whether the scan defined above can reach the region favored by the updated fit. We retain points for which $\Delta S$ and $\Delta T$ of Eq.~\eqref{eq:ST} lie inside the correlated $95\%$ confidence ellipse in the $(S,T)$ plane, and which satisfy at $95\%$ CL the applicable pair and single production limits and the Higgs coupling measurements in the $\kappa$ framework. For Higgs couplings we use the full rotations and loop functions of Sec.~\ref{sec:other}, and compare the predictions for $(\kappa_t,\kappa_W,\kappa_Z,\kappa_b,\kappa_\gamma,\kappa_g)$ with the published central values, uncertainties and observed correlations of the ATLAS effective coupling fit with free $\kappa_c$ \cite{ATLAS:2026pdi}.\footnote{For the retained sample, the zero momentum estimates of Eqs.~\eqref{eq:kF} and \eqref{eq:kVapp} differ from the full rotations by less than $1.5\%$ for $\kappa_t$, $0.001\%$ for $\kappa_b$ and $0.02\%$ for $\kappa_W$ and $\kappa_Z$, relative to the full couplings.} We require the six couplings to lie inside the approximate joint $95\%$ region obtained from a Gaussian reconstruction, with $\kappa_\tau$, $\kappa_\mu$, $\kappa_{Z\gamma}$ and $\kappa_c$ left free. This is an approximate criterion rather than a dedicated Higgs likelihood analysis.

We also checked that the residual shift in $Z\bar b_Lb_L$ is smaller than its experimental uncertainty for the retained points. As discussed in Sec.~\ref{sec:ewpt}, $S$ constrains the spin-one masses, with the lightest state at about $2.8~\mathrm{TeV}$ in the retained sample.\footnote{Limits obtained in the narrow width approximation cannot be applied directly when the large composite couplings produce broad spin-one resonances~\cite{Barducci:2015vyf,Liu:2019bua}.}  Within these constraints, the ranges above are chosen to sample the region that moves toward the direction preferred by the fit, so the point density has no probabilistic meaning.

\begin{figure}[!htbp]
  \centering
  \includegraphics[width=0.94\textwidth]{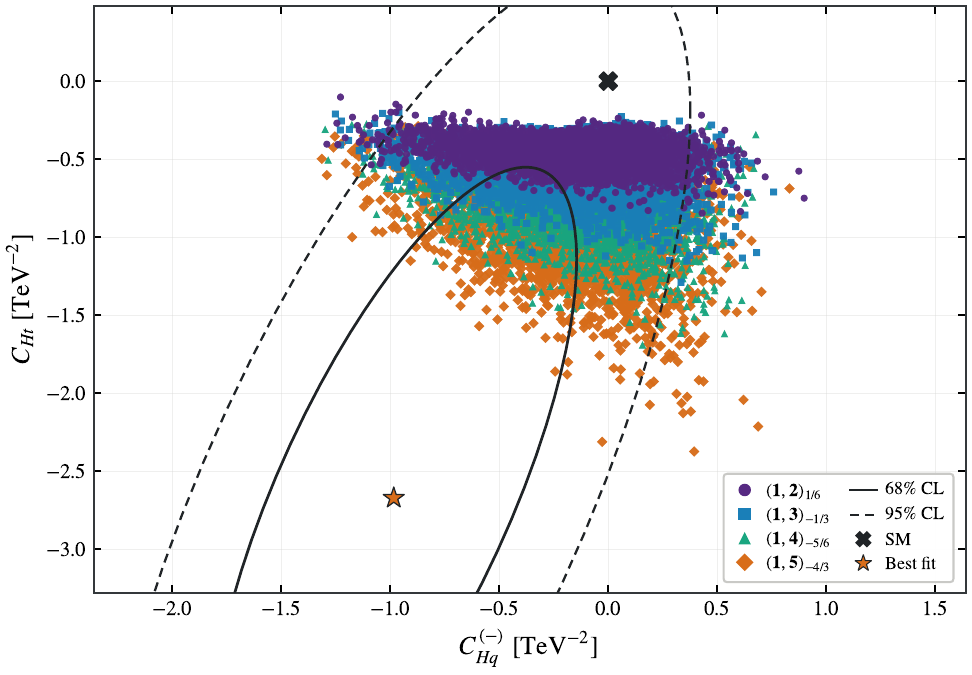}
  \caption{Predictions of the four $T_n$ embeddings in the $(C_{Hq}^{(-)},C_{Ht})$ plane for the points satisfying the top mass, oblique parameter, pair and single production and Higgs coupling constraints. Neither single production search covers bitriplet masses above $2.6~\mathrm{TeV}$. Solid and dashed contours show the $68\%$ and $95\%$ confidence regions of the updated fit. The cross and star mark the SM and the best fit point.}
  \label{fig:Cnum}
\end{figure}

Figure~\ref{fig:Cnum} shows the resulting predictions in the $(C_{Hq}^{(-)},C_{Ht})$ plane. The four embeddings reach the quadrant preferred by the fit and overlap the updated confidence region. Increasing $n$ extends the sample toward more negative $C_{Ht}$, as expected from the spin-one contribution in Eq.~\eqref{eq:CVR}, which grows with $n$, while $C_{Hq}^{(-)}$ is controlled mainly by the $q_L$ embedding, common to the four cases, and changes much less.

\subsection{Constraints from the \texorpdfstring{$T$}{T} parameter}

\begin{figure}[!htbp]
  \centering
  \includegraphics[width=0.98\textwidth]{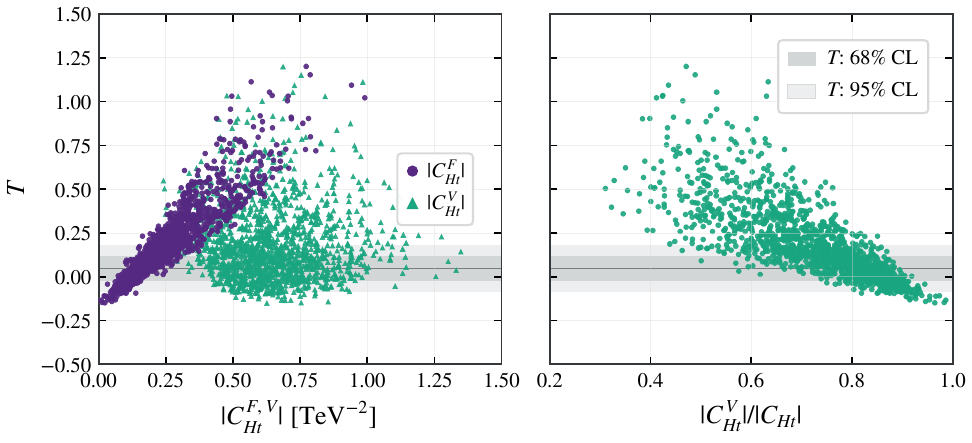}
  \caption{Fermionic and spin-one contributions to $C_{Ht}$ against $T$ for $T_n\sim(\mathbf 1,\mathbf 5)_{-4/3}$. The left panel shows $|C_{Ht}^{F}|$ (violet circles) and $|C_{Ht}^{V}|$ (green triangles), and the right panel the fraction $|C_{Ht}^{V}|/|C_{Ht}|$. Both contributions are negative. The shaded bands are the one-dimensional $68\%$ and $95\%$ intervals of Eq.~\eqref{eq:Tfit}. The points lie inside the $68\%$ region of the updated fit and satisfy the top mass, resonance mass, direct search, Higgs coupling and one-dimensional $95\%$ $S$ constraints, while $T$ is left unconstrained.}
  \label{fig:CTnum}
\end{figure}

The oblique parameter constraints remove many points that would otherwise lie inside the $68\%$ region of the updated fit. To identify the origin of the fermionic contribution to $T$, we keep the top mass, resonance mass, direct search and Higgs coupling requirements and the one-dimensional $95\%$ interval in $S$, select the points inside the $68\%$ region, and leave $T$ unconstrained. Figure~\ref{fig:CTnum} shows this sample for $T_n\sim(\mathbf 1,\mathbf 5)_{-4/3}$, where the spin-one term can be largest, and the other embeddings behave in the same way.

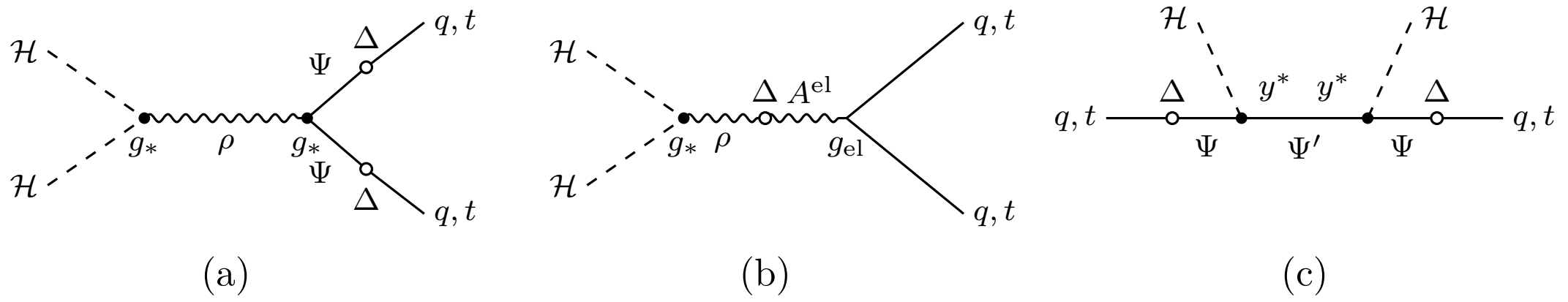
\begin{figure}[!htbp]
  \centering
  \begin{tikzpicture}[
      line width=0.6pt, scale=0.92,
      vec/.style={decorate,decoration={snake,amplitude=1pt,segment length=4.6pt}},
      vtx/.style={circle,fill,inner sep=1.05pt},
      mix/.style={circle,draw,fill=white,inner sep=1.05pt},
      lbl/.style={font=\scriptsize,inner sep=1.4pt},
      pan/.style={font=\small}
    ]
    \begin{scope}[xshift=0cm]
      \coordinate (L) at (-0.80,0); \coordinate (R) at (0.80,0);
      \draw[dashed] (-1.75, 0.66) -- (L);
      \draw[dashed] (-1.75,-0.66) -- (L);
      \draw[vec] (L) -- (R);
      \draw (R) -- (1.38, 0.50) node[mix]{} -- (1.95, 0.94);
      \draw (R) -- (1.38,-0.50) node[mix]{} -- (1.95,-0.94);
      \node[vtx] at (L) {}; \node[vtx] at (R) {};
      \node[lbl,anchor=east]  at (-1.78, 0.66) {$\mathcal H$};
      \node[lbl,anchor=east]  at (-1.78,-0.66) {$\mathcal H$};
      \node[lbl,anchor=west]  at ( 1.98, 0.94) {$q,t$};
      \node[lbl,anchor=west]  at ( 1.98,-0.94) {$q,t$};
      \node[lbl,anchor=north] at ( 0.00,-0.10) {$\rho$};
      \node[lbl,anchor=south east] at (1.12, 0.36) {$\Psi$};
      \node[lbl,anchor=north east] at (1.12,-0.36) {$\Psi$};
      \node[lbl,anchor=north] at (-0.80,-0.12) {$g_*$};
      \node[lbl,anchor=north] at ( 0.80,-0.12) {$g_*$};
      \node[lbl,anchor=south] at ( 1.38, 0.62) {$\Delta$};
      \node[lbl,anchor=north] at ( 1.38,-0.62) {$\Delta$};
      \node[pan] at (0,-1.55) {(a)};
    \end{scope}
    \begin{scope}[xshift=5.3cm]
      \coordinate (L) at (-0.80,0); \coordinate (R) at (0.80,0);
      \draw[dashed] (-1.75, 0.66) -- (L);
      \draw[dashed] (-1.75,-0.66) -- (L);
      \draw[vec] (L) -- (0,0); \draw[vec] (0,0) -- (R);
      \draw (R) -- (1.95, 0.94);
      \draw (R) -- (1.95,-0.94);
      \node[vtx] at (L) {}; \node[mix] at (0,0) {};
      \node[lbl,anchor=east]  at (-1.78, 0.66) {$\mathcal H$};
      \node[lbl,anchor=east]  at (-1.78,-0.66) {$\mathcal H$};
      \node[lbl,anchor=west]  at ( 1.98, 0.94) {$q,t$};
      \node[lbl,anchor=west]  at ( 1.98,-0.94) {$q,t$};
      \node[lbl,anchor=north] at (-0.42,-0.08) {$\rho$};
      \node[lbl,anchor=south] at ( 0.44, 0.08) {$A^{\mathrm{el}}$};
      \node[lbl,anchor=north] at (-0.80,-0.12) {$g_*$};
      \node[lbl,anchor=north] at ( 0.80,-0.12) {$g_{\mathrm{el}}$};
      \node[lbl,anchor=south] at ( 0.00, 0.10) {$\Delta$};
      \node[pan] at (0,-1.55) {(b)};
    \end{scope}
    \begin{scope}[xshift=10.6cm]
      \draw (-1.95,0) -- (-1.30,0) node[mix]{} -- (-0.62,0);
      \draw (-0.62,0) -- (0.62,0);
      \draw ( 0.62,0) -- (1.30,0) node[mix]{} -- (1.95,0);
      \draw[dashed] (-0.62,0) -- (-1.05,0.95);
      \draw[dashed] ( 0.62,0) -- ( 1.05,0.95);
      \node[vtx] at (-0.62,0) {}; \node[vtx] at (0.62,0) {};
      \node[lbl,anchor=east]  at (-1.98,0) {$q,t$};
      \node[lbl,anchor=west]  at ( 1.98,0) {$q,t$};
      \node[lbl,anchor=east]  at (-1.08,0.99) {$\mathcal H$};
      \node[lbl,anchor=west]  at ( 1.08,0.99) {$\mathcal H$};
      \node[lbl,anchor=north] at ( 0.00,-0.10) {$\Psi'$};
      \node[lbl,anchor=north] at (-0.96,-0.10) {$\Psi$};
      \node[lbl,anchor=north] at ( 0.96,-0.10) {$\Psi$};
      \node[lbl,anchor=south] at (-1.30, 0.10) {$\Delta$};
      \node[lbl,anchor=south] at ( 1.30, 0.10) {$\Delta$};
      \node[lbl,anchor=south west] at (-0.52, 0.10) {$y^*$};
      \node[lbl,anchor=south east] at ( 0.52, 0.10) {$y^*$};
      \node[pan] at (0,-1.55) {(c)};
    \end{scope}
  \end{tikzpicture}
  \caption{Schematic origin of the tree level matching contributions collected in table~\ref{tab:vlq}. Panels (a) and (b) separate heavy vector exchange through the composite fermion current from the universal contribution induced by elementary/composite gauge mixing. In (a), the vector probes the charge of the composite state $\Psi$ selected by the fermion mixing, whereas (c) shows heavy fermion exchange between two composite Yukawa vertices. The diagrams are organized as a mass insertion expansion, whose leading terms make the embedding dependence of the matching explicit.}
  \label{fig:rho}
\end{figure}

The two contributions to $C_{Ht}$ have very different effects on $T$. The spin-one term comes from the exchange of Fig.~\ref{fig:rho}(a), which respects custodial symmetry and does not contribute to $T$ at tree level, Eq.~\eqref{eq:Tv}. The negative fermionic term is dominated by the $Q'_n$ branch, which is mediated by the mixing $a_n$, responsible for breaking custodial symmetry and raising $T$ at one loop order. In Fig.~\ref{fig:CTnum}, $T$ grows approximately linearly with $|C_{Ht}^F|$, while sizable $|C_{Ht}^V|$ remains compatible with small $T$. Reaching the $68\%$ region while satisfying the $T$ constraint therefore favors $|C_{Ht}^V|$ of the same order as, or larger than, $|C_{Ht}^F|$, as shown in the right panel.

Although the Yukawa coupling $y_{Q'}^*$ is custodially invariant, it transmits the breaking from the right-handed mixing $a_n$ to the electroweak spectrum. This custodial breaking from the fermion sector is controlled by $a_n y_{Q'}^*/M_{Q'}$, whose square also determines the negative $Q'_n$ contribution to $C_{Ht}$ in Eq.~\eqref{eq:CFR}.

\FloatBarrier

\subsection{Four-quark operators}
\label{sec:4q}

The spin-one resonances also generate the four-quark operators of Eq.~\eqref{eq:4f}, which can be probed in $t\bar t t\bar t$ and $t\bar t b\bar b$ production. To confront them with data, we extend the linear fit of Sec.~\ref{sec:fit} by allowing $C_{qq}^{(1)}$, $C_{qq}^{(3)}$, $C_{qt}^{(1)}$ and $C_{tt}$ to vary. We set $C_{qt}^{(8)}=0$, as motivated by the tree level matching of the color singlet resonances included in the model.

With the observables and predictions used here, the appreciable sensitivity to these coefficients comes from the inclusive $t\bar t t\bar t$ and $t\bar t b\bar b$ cross sections, each of which fixes a single combination~\cite{Celada:2024mcf}. Four-top production probes $C_{qq}^{(1)}$ and $C_{qq}^{(3)}$ only through their sum, while $t\bar tb\bar b$ separates them~\cite{DHondt:2018cww}. The remaining directions are approximately flat. The two combinations, obtained as described in Appendix~\ref{app:4q}, are
\begin{align}
  L_1 &= 0.565\,C_{qq}^{(1)}+0.610\,C_{qq}^{(3)}-0.296\,C_{qt}^{(1)}+0.471\,C_{tt},
  \notag\\
  L_2 &= -0.427\,C_{qq}^{(1)}+0.792\,C_{qq}^{(3)}+0.248\,C_{qt}^{(1)}-0.359\,C_{tt},
  \label{eq:L12}
\end{align}
where $L_1$ is constrained mainly by $t\bar t t\bar t$ and $L_2$ by $t\bar t b\bar b$. We find $L_1=-2.05\pm1.06~\mathrm{TeV}^{-2}$ and $L_2=2.17\pm4.78~\mathrm{TeV}^{-2}$. Including these coefficients does not appreciably change the $(C_{Hq}^{(-)},C_{Ht})$ region of Fig.~\ref{fig:fit}.

The preference for negative $L_1$ reflects the four-top cross sections measured by ATLAS and CMS~\cite{ATLAS:2023ajo,CMS:2023ftu}, whose central values lie above the SM prediction, although each measurement remains compatible with it. For $C_{qq}^{(1)}$, $C_{qq}^{(3)}$ and $C_{tt}$, the interference with the electroweak SM amplitudes dominates over the QCD interference and has the opposite sign~\cite{Aoude:2022deh}, so negative values increase the four-top cross section. The SM point in the $(L_1,L_2)$ plane lies outside the joint $68\%$ region but inside the $95\%$ one.

\begin{figure}[h]
  \centering
  \includegraphics[width=0.98\textwidth]{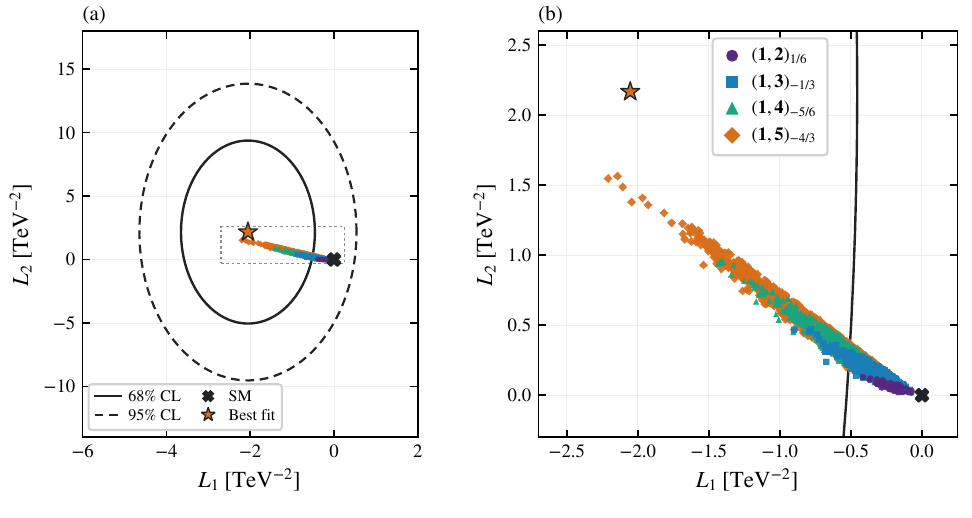}
  \caption{Linear constraints on the combinations of Eq.~\eqref{eq:L12}. Solid and dashed contours denote the joint $68\%$ and $95\%$ confidence regions. The cross and star mark the SM and the best fit point. The colored points are the model points of Fig.~\ref{fig:Cnum} inside the $68\%$ region of the top fit. Panel (b) enlarges the model region.}
  \label{fig:L12}
\end{figure}

The model gives negative $C_{qq}^{(1)}$, $C_{qq}^{(3)}$ and $C_{tt}$, Eqs.~\eqref{eq:Cqq3}--\eqref{eq:Ctt}. Figure~\ref{fig:L12} shows the model points inside the $68\%$ region of the top fit. For these points, the total four-quark contribution increases the four-top cross section by up to an amount comparable to the measured excesses. The reach of the larger representations reflects the enhanced coupling of $\rho^\perp$ to $t_R$, which increases the magnitude of $C_{tt}$. Compared with the fit of $L_1$ and $L_2$, all these points describe the data better than vanishing four-quark coefficients, none lies outside the $95\%$ region, and the fraction inside the $68\%$ region grows with the representation.

\section{Expected signals}
\label{sec:sig}
The main distinctive direct signal of the model at colliders is the presence of new VLQs, which could be produced singly or in pairs through electroweak and QCD interactions, respectively. Whereas the former is model dependent, the latter depends only on the VLQ mass. Once produced, the identification of these states depends on their width and decay modes. For scan points within the $68\%$ CL region of the updated fit, the lightest VLQ has a mass of about $1.5$--$3.1~\mathrm{TeV}$, with its charge varying across the sample.

The bitriplet $\Xi$, which provides the only negative contribution to $C_{Hq}^{(-)}$, also gives several signals through its exotic states and its effects on Higgs couplings. Its charge $-4/3$ state $Y$ has $Y\to W^-b$ as its only decay into SM particles, so $\mathrm{BR}(Y\to Wb)=1$ when decays into other resonances are closed. It is singly produced through $Wb$ fusion, with a left-handed coupling $(g/\sqrt2)\kappa_Y$ whose leading-order value is
\begin{equation}
  \kappa_Y\simeq\frac{s_q y_\Xi^*v_H}{\sqrt2 M_\Xi},
\end{equation}
which is the normalization used by the ATLAS and CMS searches~\cite{ATLAS:2025bzt,CMS:2026vwc}. At leading order, the square of this coupling is proportional to the magnitude of the bitriplet contribution to $C_{Hq}^{(-)}$. In the scan, points inside the $68\%$ region of the updated fit typically have $\kappa_Y$ between $0.2$ and $0.4$.

The CMS search gives the stronger limits but assumes narrow widths, so we apply it for $\Gamma_Y/m_Y<0.1$ within its mass range of $0.7$ to $2.4~\mathrm{TeV}$. The ATLAS search includes finite width and interference effects for $\Gamma_Y/m_Y<0.5$ and covers masses from $1.15$ to $2.6~\mathrm{TeV}$. Extending the CMS analysis to finite widths and both searches to higher masses would provide an important test of the bitriplet responsible for the negative shift in $C_{Hq}^{(-)}$.

The bitriplet also contains a charge $8/3$ state, whose decay into $tW^+W^+$ provides a complementary signature with leptons of the same sign, as studied in Ref.~\cite{Matsedonskyi:2014lla}. The bidoublet and the bitriplet also contain charge $5/3$ states, whose only decay into SM particles is $tW^+$, leading to leptons of the same sign in pair production~\cite{CMS:2024bni}. The larger representations also contain exotic charges whose decays through virtual partners can lead to final states with several $W$ bosons, providing additional signatures beyond those considered here. Single production of the charge $-1/3$ partners decaying into $tW$~\cite{CMS:2026iyo} and of the charge $2/3$ partners decaying into $Wb$~\cite{CMS:2026vwc,ATLAS:2025bzt} provides complementary channels.

The spin-one interactions that extend $C_{Ht}$ toward negative values for larger $n$ can be tested in four-top production through $pp\to t\bar t\rho\to t\bar t t\bar t$. When the vectors are produced on shell this channel can dominate, so the contact approximation can underestimate the signal~\cite{Darme:2021gtt}. This channel has also been studied as a probe of broad vector resonances using final states with two same-sign leptons~\cite{Liu:2019bua}. ATLAS has searched for this process up to $3~\mathrm{TeV}$, including broad vector benchmarks~\cite{ATLAS:2026kdx}. In most of the fit region the lightest vector is heavier than $3~\mathrm{TeV}$ and can be broad, especially for the larger representations, making extensions to higher masses and larger widths particularly relevant. Decays into fermion partner pairs offer additional channels when kinematically allowed~\cite{Barducci:2015vyf}.

There are also indirect signals of the new states. The Higgs couplings of Sec.~\ref{sec:other} provide complementary tests of the model. The mixing $s_q$ reduces $\kappa_b$ relative to the common rescaling $v/v_H$ and adds a heavy fermion contribution to $\kappa_g$, connecting Higgs measurements to single production of $Y$. Equation~\eqref{eq:kF} makes this relation explicit through $\epsilon_b$, which is proportional to $\kappa_Y^2$ at leading order. Inside the $68\%$ region of the updated fit, $\kappa_b$ lies between $0.85$ and $0.94$, $\kappa_t$ between $0.79$ and $0.92$, and $\kappa_W\simeq\kappa_Z$ between $0.88$ and $0.94$. The projected HL-LHC precisions of about $3.6\%$ for $\kappa_b$, $3.4\%$ for $\kappa_t$ and $1.6\%$ for $\kappa_W$ and $\kappa_Z$ make these couplings promising tests of the model~\cite{ATLAS:2025eii}. The loop couplings also offer useful tests, with $\kappa_g$ between $1.01$ and $1.10$ and $\kappa_\gamma$ between $0.94$ and $0.97$, compared with projected precisions of $2.4\%$ and $1.8\%$, respectively.

Boosted $tHq$ production offers a complementary probe of the electroweak top and Higgs couplings through its enhanced sensitivity to SMEFT effects~\cite{Guchait:2022jfb}. However, the small SM production rate and substantial backgrounds make differential measurements in this region experimentally challenging.

The four-quark interactions generated by the spin-one exchange provide another indirect test. A combined analysis of the CMS four-top SMEFT results~\cite{CMS:2026zfh}, which use kinematic distributions, and the differential $t\bar tb\bar b$ measurements~\cite{CMS:2023xjh,ATLAS:2024aht} would exploit the complementarity of the two processes and improve the sensitivity to the four-quark interactions of the model beyond the inclusive rates used here. The larger datasets of Run~3 and the HL-LHC will improve both measurements and could enable differential $t\bar tt\bar t$ measurements, further increasing the sensitivity~\cite{DHondt:2018cww,Aoude:2022deh}.

\section{Conclusions}
\label{sec:conc}
In this article we have analyzed the recent ATLAS and CMS measurements of $t\bar t Z$ and $tWZ$, which provide direct access to the electroweak top couplings and offer high sensitivity to BSM physics interacting with the top quark and the electroweak gauge bosons. We have considered first an analysis in the SMEFT, with dimension-six operators. By including only the operators that modify $t\bar t Z$ and $tWZ$ to leading order, we found that, together with the $b\bar bZ$ measurements, the pattern $C_{Hq}^{(-)}<0$, $C_{Ht}<0$ and $C_{Hq}^{(+)}\simeq0$ is selected. We then performed a global analysis at linear order including 46 independent WCs with SMEFiT v3.1.1, updating the public v3.1 database with the recent measurements from ATLAS, CMS, CDF and D0 of table~\ref{tab:data}. The updated fit favors $C_{Hq}^{(-)}<0$ and $C_{Ht}<0$, whereas the v3.1 database favored positive values of both. The new data shrink the allowed region and shift the best fit point, so that the SM point, which was within the $68\%$ CL region in this plane, lies at approximately $86\%$ CL.

Next, we have studied UV completions that could reproduce the selected pattern of WCs. Since the global fit favors WCs with tree level size, we have proposed a CHM with new resonances at the TeV scale, protected by custodial and $P_{LR}$ symmetry, which preserves the $Z\bar b_Lb_L$ coupling. In this construction, $C_{Hq}^{(-)}<0$ demands the presence of $SU(2)_L$ triplet VLQs, whereas $C_{Ht}<0$ requires $SU(2)_L$ singlets with positive $T_R^3$ mixing with $t_R$. Performing a numerical scan over the parameter space of the model, we have found that the pattern of WCs obtained in the global fit can be reproduced within the $68\%$ CL region. In the parameter region studied, electroweak precision constraints, through both $S$ and $T$, limit how far the model can move toward more negative $C_{Hq}^{(-)}$ and $C_{Ht}$. In particular, the $T$ parameter limits the negative fermionic contribution to $C_{Ht}$, favoring a spin-one contribution of comparable or greater magnitude in the region preferred by the updated fit. Since the spin-one contribution grows as $1/f^2$, it is in turn bounded by $S$, whose tree level contribution is enhanced at a low compositeness scale. Increasing the accuracy of the measurements of $t\bar t Z$ and $tWZ$ will test the model presented here.

The four-quark interactions provide a further test through two combinations constrained by the extended linear fit. In the plane of these two combinations, the model points inside the $68\%$ region of the updated fit describe the data better than vanishing four-quark coefficients and all remain inside the joint $95\%$ region.

The CHM presented here predicts new resonances at a few TeV. The most promising direct signals at the LHC are double and single production of exotic vectorlike quarks, with the electroweak single production of the fermion with charge $-4/3$ correlated with $C_{Hq}^{(-)}$. Another direct signal is the production of broad electroweak spin-one resonances in association with top-antitop pairs. The model also predicts indirect signals, such as modifications of order $10\%$ in $\kappa_b$ and $\kappa_t$, percent level corrections to $\kappa_g$ and $\kappa_\gamma$, and an enhanced $t\bar tt\bar t$ cross section.

The top electroweak couplings have also been analyzed recently in the SMEFT and in BSM theories with new particles in Ref.~\cite{Englert:2026vzi}.

\acknowledgments
We thank Carlos E.~M.~Wagner for useful discussions. This work has been partially supported by CONICET, project PIP-11220200101426. J.F. is supported by a doctoral fellowship from CONICET.

\appendix
\section{Details of the global fit}
\label{app:fit}

We take the Wilson coefficients to be real and absorb $1/\Lambda^2$ into their definition, quoting them in $\mathrm{TeV}^{-2}$. The flavor symmetry is $U(2)_q\times U(2)_u\times U(3)_d\times[U(1)_\ell\times U(1)_e]^3$. We use the $\{m_W,m_Z,G_F\}$ electroweak input scheme, SMEFiT v3.1.1 \cite{Giani:2023gfq}, and database v3.1 \cite{SMEFiTDatabase:v3.1}. Coefficient conventions and uncertainties follow SMEFiT3.0 \cite{Celada:2024mcf}. The v3.1 database defines the reference dataset. In the updated dataset the measurements of table~\ref{tab:data} replace earlier measurements of the same processes, and measurements with overlapping event samples are removed. The rest of the setup is common to both fits, and observables reported in a common fit retain their published joint likelihood or covariance information.

\subsection{Wilson coefficients and conventions}
\label{app:coords}

The $46$ independent Wilson coefficients of the main fit are the $45$ of the linear SMEFiT3.0 analysis~\cite{Celada:2024mcf} together with $c_\varphi$, with the same notation and definitions in terms of the Warsaw basis~\cite{Ethier:2021bye,Celada:2024mcf}. Most coincide with Warsaw basis coefficients, and the rest are linear combinations adapted to the flavor symmetry and fitted processes. In the main text we use $C_{Hq}^{(-)}=c_{\varphi Q}^{(-)}$, $C_{Hq}^{(+)}=c_{\varphi Q}^{(-)}+2c_{\varphi Q}^{(3)}$ and $C_{Ht}=c_{\varphi t}$, and the four-quark coefficients of Sec.~\ref{sec:other} are the third-generation components $C_{qq}^{(1,3)}=[C_{qq}^{(1,3)}]_{3333}$, $C_{qt}^{(1,8)}=[C_{qu}^{(1,8)}]_{3333}$ and $C_{tt}=[C_{uu}]_{3333}$.

\subsection{One-dimensional profile constraints}
\label{app:prof}

The one-dimensional profile of each fit coefficient is obtained by minimizing $\chi^2$ over the other $45$, and the resulting intervals are given in table~\ref{tab:prof}. The two profiles whose $95\%$ confidence intervals exclude zero can be traced to features already present in the input data. The preference for $c_{Qq}^{(3,1)}$ follows the positive tendency reported in the ATLAS $t$-channel single top analysis \cite{ATLAS:2025adk}, while that for $c_{\varphi d}$ was already present in the published linear SMEFiT3.0 fit~\cite{Celada:2024mcf} and may be connected to the longstanding $A_{\mathrm{FB}}^{0,b}$ tension in the $Zb\bar b$ electroweak precision data~\cite{Gori:2015nqa}.

{\footnotesize
\setlength{\tabcolsep}{6pt}
\renewcommand{\arraystretch}{0.95}
\begin{longtable}{
  @{}
  >{\raggedright\arraybackslash}p{0.28\textwidth}
  >{\centering\arraybackslash}p{0.13\textwidth}
  >{\centering\arraybackslash}p{0.25\textwidth}
  >{\centering\arraybackslash}p{0.25\textwidth}
  @{}
}
  \caption{One-dimensional profile likelihood constraints from the
  updated fit for the $46$ independent coefficients, in
  $\mathrm{TeV}^{-2}$.}
  \label{tab:prof} \\
  \toprule
  Coefficient & Best fit & $68\%$ CL & $95\%$ CL \\
  \midrule
  \endfirsthead
  \multicolumn{4}{c}{\tablename~\thetable\ (continued)} \\
  \toprule
  Coefficient & Best fit & $68\%$ CL & $95\%$ CL \\
  \midrule
  \endhead
  \midrule
  \multicolumn{4}{r}{Continued on next page} \\
  \endfoot
  \bottomrule
  \endlastfoot

  \multicolumn{4}{@{}l}{\textit{Third generation electroweak currents}} \\
  $c_{\varphi Q}^{(3)}$ & $0.458$ & $[0.181,\,0.734]$ & $[-0.084,\,1.000]$ \\
  $c_{\varphi t}$ & $-2.673$ & $[-4.077,\,-1.268]$ & $[-5.426,\,0.081]$ \\
  $c_{\varphi Q}^{(-)}$ & $-0.984$ & $[-1.539,\,-0.428]$ & $[-2.073,\,0.106]$ \\
  \addlinespace[2pt]

  \multicolumn{4}{@{}l}{\textit{Four-quark coefficients with two third-generation quarks}} \\
  $c_{Qq}^{(1,1)}$ & $0.425$ & $[-5.098,\,5.948]$ & $[-10.400,\,11.251]$ \\
  $c_{Qq}^{(3,1)}$ & $0.159$ & $[0.102,\,0.216]$ & $[0.047,\,0.270]$ \\
  $c_{td}^{(1)}$ & $-16.317$ & $[-34.689,\,2.055]$ & $[-52.325,\,19.691]$ \\
  $c_{Qd}^{(1)}$ & $7.456$ & $[-7.811,\,22.723]$ & $[-22.467,\,37.379]$ \\
  $c_{tq}^{(1)}$ & $5.913$ & $[0.749,\,11.077]$ & $[-4.209,\,16.034]$ \\
  $c_{Qu}^{(1)}$ & $-12.014$ & $[-19.563,\,-4.465]$ & $[-26.810,\,2.782]$ \\
  $c_{tu}^{(1)}$ & $6.177$ & $[-4.135,\,16.489]$ & $[-14.034,\,26.388]$ \\
  $c_{Qq}^{(1,8)}$ & $-1.878$ & $[-3.268,\,-0.487]$ & $[-4.603,\,0.848]$ \\
  $c_{Qq}^{(3,8)}$ & $-0.838$ & $[-3.010,\,1.335]$ & $[-5.096,\,3.420]$ \\
  $c_{td}^{(8)}$ & $4.810$ & $[-0.694,\,10.314]$ & $[-5.977,\,15.598]$ \\
  $c_{Qd}^{(8)}$ & $2.927$ & $[-1.421,\,7.274]$ & $[-5.595,\,11.448]$ \\
  $c_{tq}^{(8)}$ & $2.125$ & $[0.640,\,3.609]$ & $[-0.786,\,5.035]$ \\
  $c_{Qu}^{(8)}$ & $-4.993$ & $[-7.758,\,-2.227]$ & $[-10.412,\,0.427]$ \\
  $c_{tu}^{(8)}$ & $1.394$ & $[-2.791,\,5.580]$ & $[-6.808,\,9.597]$ \\
  \addlinespace[2pt]

  \multicolumn{4}{@{}l}{\textit{Lepton currents}} \\
  $c_{\varphi\ell_1}^{(1)}$ & $0.059$ & $[-0.082,\,0.199]$ & $[-0.217,\,0.335]$ \\
  $c_{\varphi\ell_1}^{(3)}$ & $0.020$ & $[-0.044,\,0.083]$ & $[-0.104,\,0.143]$ \\
  $c_{\varphi\ell_2}^{(1)}$ & $0.006$ & $[-0.159,\,0.172]$ & $[-0.318,\,0.331]$ \\
  $c_{\varphi\ell_2}^{(3)}$ & $0.062$ & $[-0.030,\,0.153]$ & $[-0.117,\,0.240]$ \\
  $c_{\varphi\ell_3}^{(1)}$ & $0.055$ & $[-0.108,\,0.219]$ & $[-0.266,\,0.376]$ \\
  $c_{\varphi\ell_3}^{(3)}$ & $0.006$ & $[-0.090,\,0.102]$ & $[-0.182,\,0.194]$ \\
  $c_{\varphi e}$ & $0.078$ & $[-0.206,\,0.363]$ & $[-0.480,\,0.636]$ \\
  $c_{\varphi\mu}$ & $0.082$ & $[-0.204,\,0.367]$ & $[-0.478,\,0.641]$ \\
  $c_{\varphi\tau}$ & $0.063$ & $[-0.222,\,0.348]$ & $[-0.495,\,0.621]$ \\
  \addlinespace[2pt]

  \multicolumn{4}{@{}l}{\textit{Light quark currents and four-lepton coefficient}} \\
  $c_{\varphi q}^{(-)}$ & $-0.006$ & $[-0.124,\,0.113]$ & $[-0.237,\,0.226]$ \\
  $c_{\varphi q}^{(3)}$ & $-0.038$ & $[-0.082,\,0.007]$ & $[-0.125,\,0.049]$ \\
  $c_{\varphi u}$ & $0.027$ & $[-0.189,\,0.243]$ & $[-0.396,\,0.450]$ \\
  $c_{\varphi d}$ & $-0.579$ & $[-0.861,\,-0.298]$ & $[-1.130,\,-0.028]$ \\
  $c_{\ell\ell}^{1221}=c_{\ell\ell}^{1111}$
    & $0.016$ & $[-0.042,\,0.075]$ & $[-0.099,\,0.132]$ \\
  \addlinespace[2pt]

  \multicolumn{4}{@{}l}{\textit{Bosonic coefficients}} \\
  $c_{\varphi}$ & $-0.832$ & $[-3.874,\,2.210]$ & $[-6.795,\,5.131]$ \\
  $c_{\varphi B}$ & $-0.049$ & $[-0.246,\,0.147]$ & $[-0.434,\,0.336]$ \\
  $c_{\varphi\Box}$ & $0.152$ & $[-0.772,\,1.076]$ & $[-1.660,\,1.963]$ \\
  $c_{\varphi D}$ & $-0.165$ & $[-0.733,\,0.403]$ & $[-1.277,\,0.947]$ \\
  $c_{\varphi G}$ & $-0.004$ & $[-0.013,\,0.005]$ & $[-0.022,\,0.014]$ \\
  $c_{\varphi W}$ & $0.106$ & $[-0.150,\,0.361]$ & $[-0.395,\,0.606]$ \\
  $c_{\varphi WB}$ & $0.054$ & $[-0.210,\,0.318]$ & $[-0.464,\,0.572]$ \\
  $c_{WWW}$ & $0.115$ & $[-0.184,\,0.415]$ & $[-0.472,\,0.703]$ \\
  \addlinespace[2pt]

  \multicolumn{4}{@{}l}{\textit{Dipole and Yukawa coefficients}} \\
  $c_{tG}$ & $0.034$ & $[-0.047,\,0.115]$ & $[-0.125,\,0.193]$ \\
  $c_{tW}$ & $-0.077$ & $[-0.133,\,-0.020]$ & $[-0.188,\,0.035]$ \\
  $c_{tZ}$ & $-0.971$ & $[-4.681,\,2.739]$ & $[-8.242,\,6.301]$ \\
  $c_{t\varphi}$ & $-0.359$ & $[-2.156,\,1.438]$ & $[-3.881,\,3.163]$ \\
  $c_{b\varphi}$ & $0.025$ & $[-0.009,\,0.058]$ & $[-0.041,\,0.091]$ \\
  $c_{c\varphi}$ & $-0.033$ & $[-0.135,\,0.068]$ & $[-0.232,\,0.165]$ \\
  $c_{\tau\varphi}$ & $0.007$ & $[-0.010,\,0.024]$ & $[-0.027,\,0.040]$ \\
\end{longtable}
}

\subsection{Extended fit with four-heavy-quark coefficients}
\label{app:4q}

The extended fit of Sec.~\ref{sec:4q} adds the coefficients $C_{qq}^{(1)}$, $C_{qq}^{(3)}$, $C_{qt}^{(1)}$ and $C_{tt}$ of Sec. ~\ref{app:coords} to the $46$ coefficients of the main fit, with $C_{qt}^{(8)}=0$. For $t\bar t t\bar t$, their linear responses include the interference with the electroweak SM amplitudes~\cite{Aoude:2022deh}, using the \texttt{LO\_QCD\_EW} tables of Ref.~\cite{SMEFiTDatabase:PR160}. The $t\bar t b\bar b$ predictions and all other inputs and uncertainties are kept as in the main fit.

After profiling the $46$ coefficients of the main fit, we diagonalize the Fisher matrix of these four coefficients. Its two orthonormal eigenvectors with appreciable sensitivity give the combinations of Eq.~\eqref{eq:L12}, whose one-dimensional intervals are given in table~\ref{tab:L12}. The other two combinations have uncertainties above $10^4~\mathrm{TeV}^{-2}$ and are fixed to zero. The joint regions of Fig.~\ref{fig:L12} use the same two-parameter thresholds as Fig.~\ref{fig:fit}. Including these combinations does not appreciably change the constraints of the main fit.

\begin{table}[htbp]
  \centering
  \begin{tabular}{@{}lccc@{}}
    \toprule
    Combination & Best fit & $68\%$ CL & $95\%$ CL \\
    \midrule
    $L_1$ & $-2.05$ & $[-3.11,\,-1.00]$ & $[-4.13,\,0.02]$ \\
    $L_2$ & $2.17$ & $[-2.61,\,6.94]$ & $[-7.19,\,11.53]$ \\
    \bottomrule
  \end{tabular}
  \caption{One-dimensional constraints on the combinations of Eq.~\eqref{eq:L12}, in $\mathrm{TeV}^{-2}$.}
  \label{tab:L12}
\end{table}

\section{Fermion bases and electroweak observables}
\label{app:ferm}

The fermion fields are written in three bases in this work. The two-site basis makes the elementary and composite content explicit. Removing the elementary/composite mixings before electroweak symmetry breaking gives the basis used for the SMEFT matching, while diagonalizing the mass matrices after electroweak symmetry breaking gives the physical spectrum required for the oblique corrections. The sequence is
\begin{equation}
  \text{two-site basis}
  \xrightarrow[v_H=0]{\text{mixings }\Delta}
  \text{electroweak symmetric basis}
  \xrightarrow[v_H\neq0]{\text{Yukawa mixings}}
  \text{physical mass basis}.
\end{equation}
We illustrate both changes of basis directly with the fermions of the model. The construction is the usual one in two-site descriptions of partial compositeness~\cite{Contino:2006nn,DeCurtis:2011yx}.

\subsection{The left-handed doublet}

Consider first the mixing of the elementary doublet with the $Y=1/6$ doublet of $Q$. Writing $Q_L\equiv Q_{1/6,L}$, the mass terms before electroweak symmetry breaking are
\begin{equation}
  \mathcal L_q^{(0)}
  =\Delta_q\,\overline q_L^{\rm el}Q_R
   -M_Q\,\overline Q_LQ_R+\text{h.c.}
\end{equation}
In the convention $\mathcal L\supset-\bar\psi_L\mathcal M\psi_R+\text{h.c.}$, with left-handed basis $(q_L^{\rm el},Q_L)$ and right-handed basis $(Q_R)$, the rectangular mass matrix is
\begin{equation}
  \mathcal M_q^{\rm site}
  =
  \begin{pmatrix}
    -\Delta_q\\ M_Q
  \end{pmatrix}.
\end{equation}
We denote the rotation to the electroweak symmetric basis by $R_q$,
\begin{equation}
  \begin{pmatrix}q_L^{(0)}\\ Q_L^{(0)}\end{pmatrix}
  =R_q\begin{pmatrix}q_L^{\rm el}\\ Q_L\end{pmatrix},
  \qquad
  R_q=\begin{pmatrix}c_q&s_q\\ -s_q&c_q\end{pmatrix},
  \qquad
  s_q=\frac{\Delta_q}{\sqrt{M_Q^2+\Delta_q^2}},
  \quad
  c_q=\frac{M_Q}{\sqrt{M_Q^2+\Delta_q^2}}.
\end{equation}
Applying this rotation to the mass matrix gives
\begin{equation}
  \mathcal M_q^{(0)}
  =
  \begin{pmatrix}
    0\\ \sqrt{M_Q^2+\Delta_q^2}
  \end{pmatrix}
  =
  \begin{pmatrix}
    0\\ M_Q/c_q
  \end{pmatrix}.
\end{equation}
In this basis $q_L^{(0)}$ is massless, whereas $Q_L^{(0)}$ pairs with $Q_R$ with mass $m_Q=M_Q/c_q$. The rotation removes the mass mixing between light and heavy states without replacing the Higgs field by its vacuum expectation value. The Yukawa interactions therefore retain vertices with one light and one heavy fermion, which enter the tree-level SMEFT matching.

\subsection{The right-handed top}

For $t_R^{\rm el}$ the same step involves the charge $2/3$ states of two composite multiplets, which we denote by $T_1$ and $T_n$. The mass terms are
\begin{equation}
  \mathcal L_t^{(0)}
  =\overline T_{1L}\bigl(\Delta_{R1}t_R^{\rm el}-M_{T_1}T_{1R}\bigr)
   +\overline T_{nL}\bigl(\Delta_{Rn}t_R^{\rm el}-M_{T_n}T_{nR}\bigr)
   +\text{h.c.}
\end{equation}
In the bases $(T_{1L},T_{nL})$ and $(t_R^{\rm el},T_{1R},T_{nR})$, this is
\begin{equation}
  \mathcal M_t^{\rm site}
  =
  \begin{pmatrix}
    -\Delta_{R1}&M_{T_1}&0\\
    -\Delta_{Rn}&0&M_{T_n}
  \end{pmatrix}.
\end{equation}
We parametrize the two mixings by a total compositeness angle $\chi$ and a direction angle $\phi$,
\begin{equation}
  \frac{\Delta_{R1}}{M_{T_1}}=\tan\chi\cos\phi,
  \qquad
  \frac{\Delta_{Rn}}{M_{T_n}}=\tan\chi\sin\phi.
\end{equation}
Here $\chi$ measures the total composite fraction of the light state, whereas $\phi$ selects its direction in the $(T_1,T_n)$ plane. This becomes explicit by first defining the parallel and orthogonal composite combinations,
\begin{equation}
  T_{\parallel R}=\cos\phi\,T_{1R}+\sin\phi\,T_{nR},
  \qquad
  T_{\perp R}=-\sin\phi\,T_{1R}+\cos\phi\,T_{nR}.
\end{equation}
The massless direction lies in the $(t_R^{\rm el},T_{\parallel R})$ plane, so the second rotation is
\begin{align}
  t_R^{(0)}&=\cos\chi\,t_R^{\rm el}+\sin\chi\,T_{\parallel R},
  &
  \widetilde T_R&=-\sin\chi\,t_R^{\rm el}+\cos\chi\,T_{\parallel R}.
\end{align}
Together these two steps define the rotation $R_t$,
\begin{equation}
  \begin{pmatrix}t_R^{(0)}\\ \widetilde T_R\\ T_{\perp R}\end{pmatrix}
  =R_t\begin{pmatrix}t_R^{\rm el}\\ T_{1R}\\ T_{nR}\end{pmatrix},
  \qquad
  R_t=
  \begin{pmatrix}
    c_\chi&s_\chi c_\phi&s_\chi s_\phi\\
    -s_\chi& c_\chi c_\phi& c_\chi s_\phi\\
    0&-s_\phi&c_\phi
  \end{pmatrix},
\end{equation}
where $s_\chi=\sin\chi$, $c_\chi=\cos\chi$ and similarly for $\phi$. Applying $R_t$ to the mass matrix gives
\begin{equation}
  \mathcal M_t^{(0)}
  =
  \begin{pmatrix}
    0&\dfrac{M_{T_1}\cos\phi}{\cos\chi}&-M_{T_1}\sin\phi\\[6pt]
    0&\dfrac{M_{T_n}\sin\phi}{\cos\chi}& M_{T_n}\cos\phi
  \end{pmatrix}.
\end{equation}
The vanishing first column identifies $t_R^{(0)}$ as the light state, with total composite fraction $\sin^2\chi$, distributed as $\sin^2\chi\cos^2\phi$ and $\sin^2\chi\sin^2\phi$ between the two resonances. The heavy $2\times2$ block is generally non-diagonal when both mixings are present. This does not affect the SMEFT matching, which can be performed with the full heavy propagator. A further rotation restricted to the heavy subspace gives the electroweak symmetric heavy mass eigenstates without changing $t_R^{(0)}$.

The two composite amplitudes of the light state are
\begin{equation}
  a_1\equiv s_\chi c_\phi,
  \qquad
  a_n\equiv s_\chi s_\phi,
  \qquad
  a_1^2+a_n^2=s_\chi^2.
\end{equation}

\subsection{From the symmetric to the physical basis}
\label{app:fphys}

For illustration, we write the full charge $2/3$ mass matrix in the two-site basis. With the mass convention used above, the field ordering is
\begin{equation}
  \Psi_{L,R}^{2/3,\rm site}=
    \bigl(t^{{\rm el},2/3},Q_{1/6}^{2/3},Q_{7/6}^{2/3},T_1^{2/3},T_n^{2/3},
  \Xi_1^{2/3},\Xi_2^{2/3},\Xi_3^{2/3},Q_{n,1/6}^{\prime\,2/3}\bigr)_{L,R}^{T},
\end{equation}
where the superscript denotes the electric charge of the component and the subscripts of $Q$ and $Q'_n$ its hypercharge. The three components of the bitriplet, $\Xi_{1,2,3}^{2/3}$, have $T_L^3=1,0,-1$, respectively, and the elementary entries denote $t_L^{{\rm el},2/3}$ and $t_R^{{\rm el},2/3}$. The mass matrix is \begingroup \small
\setlength{\arraycolsep}{3pt}
\renewcommand{\arraystretch}{1.25}
\begin{equation}
  \mathcal M_{2/3}^{\rm site}(v_H)=
  \begin{pmatrix}
  0&-\Delta_q&0&0&0&0&0&0&0\\
  0&M_Q&0&\frac{y_t^*v_H}{\sqrt2}&0&-\frac{y_\Xi^*v_H}{\sqrt2}&\frac{y_\Xi^*v_H}{2\sqrt2}&0&0\\
  0&0&M_Q&\frac{y_t^*v_H}{\sqrt2}&0&0&\frac{y_\Xi^*v_H}{2\sqrt2}&-\frac{y_\Xi^*v_H}{\sqrt2}&0\\
  -\Delta_{R1}&0&0&M_{T_1}&0&0&0&0&0\\
  -\Delta_{Rn}&0&0&0&M_{T_n}&0&0&0&0\\
  0&0&0&0&0&M_\Xi&0&0&0\\
  0&0&0&0&0&0&M_\Xi&0&0\\
  0&0&0&0&0&0&0&M_\Xi&0\\
  0&0&0&0&\frac{y_{Q'}^*v_H}{\sqrt2}&0&0&0&M_{Q'}
  \end{pmatrix}.
\end{equation}
\endgroup The phase of $Q'_n$ is chosen so that its Yukawa entry has the sign shown. The symmetric basis matrix is $\mathcal M_{2/3}^{(0)}=R_L\mathcal M_{2/3}^{\rm site}R_R^\dagger$, where $R_L$ and $R_R$ embed the rotations $R_q$ and $R_t$ above. In either basis the Higgs coupling matrix is the derivative with respect to $v_H$, with the other parameters held fixed.

After electroweak symmetry breaking, the Yukawa interactions generate further mixings among fields with the same electric charge. For each charge sector $\mathcal Q$, the matrix in the electroweak symmetric basis changes as
\begin{equation}
  \mathcal M_{\mathcal Q}^{(0)}(0)
  \longrightarrow
  \mathcal M_{\mathcal Q}^{(0)}(v_H)
  =\mathcal M_{\mathcal Q}^{(0)}(0)+\delta\mathcal M_{\mathcal Q}(v_H),
\end{equation}
where $\delta\mathcal M_{\mathcal Q}(v_H)$ contains the Yukawa entries induced by the Higgs vacuum expectation value. The unitary matrices $U_L^{\mathcal Q}$ and $U_R^{\mathcal Q}$ diagonalize the full mass matrix and define the physical fields through
\begin{equation}
  \Psi_\eta^{\mathcal Q,(0)}=U_\eta^{\mathcal Q}\psi_\eta^{\mathcal Q},
  \qquad
  U_L^{\mathcal Q\dagger}\mathcal M_{\mathcal Q}^{(0)}(v_H)U_R^{\mathcal Q}
  =\widehat{\mathcal M}_{\mathcal Q}
  =\operatorname{diag}(m_1^{\mathcal Q},\ldots,m_{N_{\mathcal Q}}^{\mathcal Q}),
  \qquad \eta=L,R.
\end{equation}
Thus $\psi_i^{\mathcal Q}$ is the mass eigenstate with mass $m_i^{\mathcal Q}$. Let $T_\eta^{3,\mathcal Q}$ and $T_\eta^{+,\mathcal Q}$ denote the weak generators in the electroweak symmetric basis, with $T_\eta^{+,\mathcal Q}$ mapping the charge $\mathcal Q-1$ sector into the charge $\mathcal Q$ sector. The neutral and charged coupling matrices in the physical basis are then
\begin{equation}
  \mathcal X_\eta^{\mathcal Q}=2U_\eta^{\mathcal Q\dagger}T_\eta^{3,\mathcal Q}U_\eta^{\mathcal Q},
  \qquad
  \mathcal V_\eta^{\mathcal Q}=U_\eta^{\mathcal Q\dagger}T_\eta^{+,\mathcal Q}U_\eta^{\mathcal Q-1},
  \qquad \eta=L,R.
  \label{eq:gphys}
\end{equation}
Together with the masses $m_i^{\mathcal Q}$, these matrices determine the ingredients needed to compute the fermionic contribution to the electroweak vacuum polarizations.

\subsection{Fermionic contributions to the \texorpdfstring{$S$ and $T$}{S and T} parameters}
\label{app:fST}

After electroweak symmetry breaking, the fermionic oblique corrections are fixed by the physical masses $m_i^{\mathcal Q}$ and by the weak current matrices of Eq.~\eqref{eq:gphys}. In the mass eigenstate notation of Ref.~\cite{Anastasiou:2009rv}, the relevant interactions are
\begin{align}
  \mathcal L_W={}&\frac{g}{\sqrt2}\,
  \overline\psi_i^{\mathcal Q}\gamma^\mu
  \left[(\mathcal V_L^{\mathcal Q})_{ij}P_L
       +(\mathcal V_R^{\mathcal Q})_{ij}P_R\right]
  \psi_j^{\mathcal Q-1}W_\mu^+ +\text{h.c.},\\
  \mathcal L_Z={}&\frac{g}{2c_W}\,
  \overline\psi_i^{\mathcal Q}\gamma^\mu
  \left[(\mathcal X_L^{\mathcal Q})_{ij}P_L
       +(\mathcal X_R^{\mathcal Q})_{ij}P_R
       -2s_W^2\mathcal Q\,\delta_{ij}\right]
  \psi_j^{\mathcal Q} Z_\mu .
\end{align}

With $s_W=\sin\theta_W$ and $y_i^{\mathcal Q}=(m_i^{\mathcal Q}/m_Z)^2$, the contribution to $T$ is
\begin{align}
  T_F^{\rm model}={}&\frac{N_c}{16\pi s_W^2c_W^2}
  \sum_{\mathcal Q,i,j}\Bigg\{
  \left(|\mathcal V_{L,ij}^{\mathcal Q}|^2+|\mathcal V_{R,ij}^{\mathcal Q}|^2\right)
  \theta_+(y_i^{\mathcal Q},y_j^{\mathcal Q-1}) \notag\\
  &\quad +2\operatorname{Re}\!\left(
  \mathcal V_{L,ij}^{\mathcal Q}\mathcal V_{R,ij}^{\mathcal Q*}\right)
  \theta_-(y_i^{\mathcal Q},y_j^{\mathcal Q-1}) \notag\\
  &\quad
  -\frac12\Bigl[
  \left(|\mathcal X_{L,ij}^{\mathcal Q}|^2+|\mathcal X_{R,ij}^{\mathcal Q}|^2\right)
  \theta_+(y_i^{\mathcal Q},y_j^{\mathcal Q}) \notag\\
  &\qquad +2\operatorname{Re}\!\left(
  \mathcal X_{L,ij}^{\mathcal Q}\mathcal X_{R,ij}^{\mathcal Q*}\right)
  \theta_-(y_i^{\mathcal Q},y_j^{\mathcal Q})\Bigr]\Bigg\},
\end{align}
where $N_c=3$. After cancellation of the regulator dependent terms in the complete charged and neutral current sums, the finite loop functions are
\begin{align}
  \theta_+(y_1,y_2)&=y_1+y_2
  -\frac{2y_1y_2}{y_1-y_2}\log\frac{y_1}{y_2},\\
  \theta_-(y_1,y_2)&=2\sqrt{y_1y_2}
  \left[\frac{y_1+y_2}{y_1-y_2}\log\frac{y_1}{y_2}-2\right].
\end{align}

The contribution to $S$ takes a similar form in the same basis~\cite{Carena:2007ua}. It is convenient to introduce the standard hypercharge matrices
\begin{equation}
  (\mathcal Y_{L,R}^{\mathcal Q})_{ij}
  =\mathcal Q\,\delta_{ij}-\frac12(\mathcal X_{L,R}^{\mathcal Q})_{ij},
\end{equation}
for which
\begin{align}
  S_F^{\rm model}={}&\frac{N_c}{2\pi}\sum_{\mathcal Q,i,j}\Bigg\{
  \mathcal X_{L,ij}^{\mathcal Q}\left[
  \mathcal Y_{L,ji}^{\mathcal Q}\bar\chi_+(y_i^{\mathcal Q},y_j^{\mathcal Q})
  +\mathcal Y_{R,ji}^{\mathcal Q}\bar\chi_-(y_i^{\mathcal Q},y_j^{\mathcal Q})\right]
  \notag\\
  &\phantom{\frac{N_c}{2\pi}\sum_{\mathcal Q,i,j}\Bigg\{}
  \llap{${}+{}$}\mathcal X_{R,ij}^{\mathcal Q}\left[
  \mathcal Y_{R,ji}^{\mathcal Q}\bar\chi_+(y_i^{\mathcal Q},y_j^{\mathcal Q})
  +\mathcal Y_{L,ji}^{\mathcal Q}\bar\chi_-(y_i^{\mathcal Q},y_j^{\mathcal Q})\right]\Bigg\}.
\end{align}
The loop functions in the same dimensionless variables are
\begin{align}
  \bar\chi_+(y_1,y_2)={}&
  \frac{5(y_1^2+y_2^2)-22y_1y_2}{9(y_1-y_2)^2} \notag\\
  &+\frac{3y_1y_2(y_1+y_2)-(y_1^3+y_2^3)}
  {3(y_1-y_2)^3}\log\frac{y_1}{y_2}
  -\frac13\log(y_1y_2),\\
  \bar\chi_-(y_1,y_2)={}&
  \frac{\sqrt{y_1y_2}}{(y_1-y_2)^3}
  \left[y_1^2-y_2^2-2y_1y_2\log\frac{y_1}{y_2}\right].
\end{align}
The scale dependence of the individual loop functions cancels in the sum over the complete fermion spectrum. The shifts entering the electroweak fit are
\begin{equation}
  \Delta S_F^{1\ell}=S_F^{\rm model}-S_F^{\rm SM},
  \qquad
  \Delta T_F^{1\ell}=T_F^{\rm model}-T_F^{\rm SM},
\end{equation}
where the SM terms are evaluated with the same physical top and bottom masses.

\section{Spin-one bases and electroweak observables}
\label{app:vec}

The spin-one fields are also written in three bases. The two-site basis keeps the elementary and composite gauge groups manifest. Diagonalizing the link mixing before electroweak symmetry breaking gives the electroweak symmetric basis, with light and heavy combinations, which is used for the tree-level SMEFT matching, while diagonalizing the mass matrices after electroweak symmetry breaking gives the physical mass eigenstates required for the oblique corrections. The sequence is
\begin{equation}
  \text{two-site basis}
  \!\xrightarrow[v_H=0]{\text{link mixing}}
  \text{electroweak symmetric basis}
  \xrightarrow[v_H\neq0]{\text{Higgs kinetic term}}
  \text{physical mass basis}.
\end{equation}
The two changes of basis are given in Appendices~\ref{app:vsym} and \ref{app:vphys}, and the spin-one contributions to the oblique parameters in Appendix~\ref{app:vST}.

\subsection{From the two-site to the electroweak symmetric basis}
\label{app:vsym}

A link field $\Omega$ transforms as $\Omega\to U_{\mathrm{el}}\Omega U_{\mathrm{cp}}^\dagger$. Its kinetic term, denoted by $|D_\mu\Omega|^2$ in Eq.~\eqref{eq:Lmix}, is
\begin{equation}
  \mathcal L_{\mathrm{link}}=\frac{f^2}{4}
  \operatorname{Tr}\!\left[(D_\mu\Omega)^\dagger D^\mu\Omega\right],
  \qquad
  D_\mu\Omega=\partial_\mu\Omega
  -ig_{\mathrm{el}}A_\mu^{\mathrm{el}}\Omega
  +ig_{\mathrm{cp}}\Omega A_\mu^{\mathrm{cp}}.
\end{equation}
We use the normalization $\operatorname{Tr}[T^AT^B]=\delta^{AB}$. This is the standard link construction used in deconstructed composite Higgs models~\cite{DeCurtis:2011yx}. For the elementary $SU(2)_L\times U(1)_Y$ and composite $SU(2)_L\times SU(2)_R\times U(1)_X$ groups, and with the choice $g_X^*=g_R^*$ of Sec.~\ref{sec:model}, it is convenient to separate the neutral composite fields into the hypercharge direction and its orthogonal combination,
\begin{equation}
  \rho_\mu^\parallel=\frac{\rho_{R\mu}^3+\rho_{X\mu}}{\sqrt2},
  \qquad
  \rho_\mu^\perp=\frac{\rho_{R\mu}^3-\rho_{X\mu}}{\sqrt2}.
\end{equation}
The field $\rho^\parallel$ is parallel to hypercharge, whereas $\rho^\perp$, like the charged states $\rho_{R\mu}^\pm=(\rho_{R\mu}^1\mp i\rho_{R\mu}^2)/\sqrt2$, has no elementary partner. In unitary gauge the link terms are
\begin{align}
  \mathcal L_{\mathrm{link}}={}&
  \frac{f^2}{4}\sum_{a=1}^{3}
    (\widehat g\widehat W_\mu^a-g_L^*\rho_{L\mu}^a)^2
  +\frac{f^2}{4}
    \left[\sqrt2\widehat g'\widehat B_\mu
    -g_R^*\rho_\mu^\parallel\right]^2
  \nonumber\\
  &+\frac{f^2g_R^{*2}}{2}\rho_{R\mu}^+\rho_R^{-\mu}
  +\frac{f^2g_R^{*2}}{4}\rho_\mu^\perp\rho^{\perp\mu}.
\end{align}
The vector Lagrangian is completed by
\begin{align}
  \mathcal L_V={}&
  -\frac14\widehat W_{\mu\nu}^a\widehat W^{a\mu\nu}
  -\frac14\widehat B_{\mu\nu}\widehat B^{\mu\nu}
  -\frac14\rho_{L\mu\nu}^a\rho_L^{a\mu\nu}
  \nonumber\\
  &-\frac14\rho_{R\mu\nu}^a\rho_R^{a\mu\nu}
  -\frac14\rho_{X\mu\nu}\rho_X^{\mu\nu}
  +\mathcal L_{\mathrm{link}}.
  \label{eq:Lv}
\end{align}

Before electroweak symmetry breaking, the fields parallel to the elementary gauge factors are diagonalized by
\begin{align}
  W_\mu^a&=c_L\widehat W_\mu^a+s_L\rho_{L\mu}^a,
  &W_\mu^{*a}&=-s_L\widehat W_\mu^a+c_L\rho_{L\mu}^a,
  \notag\\
  B_\mu&=c_Y\widehat B_\mu
    +s_Y\rho_\mu^\parallel,
  &B_\mu^*&=-s_Y\widehat B_\mu
    +c_Y\rho_\mu^\parallel.
  \label{eq:vrot}
\end{align}
The low energy couplings obey $g=\widehat g c_L=g_L^*s_L$ and $g'=\widehat g'c_Y=(g_R^*/\sqrt2)s_Y$, and therefore
\begin{equation}
  \frac1{g^2}=\frac1{\widehat g^2}+\frac1{g_L^{*2}},
  \qquad
  \frac1{g'^2}=\frac1{\widehat g'^2}+\frac2{g_R^{*2}},
\end{equation}
while the heavy masses are those of Eq.~\eqref{eq:mV}.

For a generator shared by the two sites, with light and heavy combinations $V_\mu$ and $V_\mu^*$ and mixing angle $\theta$ ($\theta_L$ or $\theta_Y$, with $\sin\theta_{L,Y}=s_{L,Y}$), the same rotation acts on its currents according to~\cite{Contino:2006nn}
\begin{align}
  g_{\mathrm{el}}A_\mu^{\mathrm{el}}J_{\mathrm{el}}^\mu
  +g_{\mathrm{cp}}A_\mu^{\mathrm{cp}}J_{\mathrm{cp}}^\mu
  ={}&gV_\mu(J_{\mathrm{el}}^\mu+J_{\mathrm{cp}}^\mu)
  \nonumber\\
  &+gV_\mu^*\left(-\tan\theta\,J_{\mathrm{el}}^\mu
  +\cot\theta\,J_{\mathrm{cp}}^\mu\right).
  \label{eq:jrot}
\end{align}
For hypercharge, $g_X^*=g_R^*$ implies that $\rho^\parallel$ couples to the composite hypercharge current with strength $g_R^*/\sqrt2$, which plays the role of $g_{\mathrm{cp}}$. The light fermion projections are fixed by the rotations in Appendix~\ref{app:ferm}. It is convenient to define
\begin{align}
  J_H^{a\mu}&=i(H^\dagger\overleftrightarrow{D}^\mu\tau^aH),
  &J_q^{a\mu}&=\bar q_L\tau^a\gamma^\mu q_L,
  \notag\\
  J_H^\mu&=i(H^\dagger\overleftrightarrow{D}^\mu H),
  &J_q^\mu&=\bar q_L\gamma^\mu q_L,
  &J_t^\mu&=\bar t_R\gamma^\mu t_R,
\end{align}
and the three combinations
\begin{equation}
  \xi_q^L=\frac{s_q^2-s_L^2}{s_Lc_L},
  \qquad
  \xi_q^Y=\frac{s_q^2-s_Y^2}{s_Yc_Y},
  \qquad
  \xi_t^Y=\frac{s_\chi^2-s_Y^2}{s_Yc_Y}.
\end{equation}
The terms linear in a heavy vector and bilinear in light fields that enter the coefficients considered here are then
\begin{align}
  \mathcal L_V^{\rm lin}\supset{}&
  \frac g2 W_\mu^{*a}
  \left(\frac{c_L}{s_L}J_H^{a\mu}+\xi_q^LJ_q^{a\mu}\right)
  \notag\\
  &+g'B_\mu^*
  \left(\frac{c_Y}{2s_Y}J_H^\mu
  +\frac{\xi_q^Y}{6}J_q^\mu
  +\frac{2\xi_t^Y}{3}J_t^\mu\right)
  \notag\\
  &+\frac{g_R^*}{\sqrt2}\rho_\mu^\perp
  \left[\frac12J_H^\mu-\frac76s_q^2J_q^\mu
  +\left(-\frac23a_1^2+\left(n-\frac53\right)a_n^2\right)J_t^\mu\right].
  \label{eq:gV}
\end{align}
The first two lines follow from the rotation of shared currents in Eq.~\eqref{eq:jrot}. The last line is instead the direct projection of the composite $T_R^3-X$ current, since $\rho^\perp$ has no elementary partner. The charged states $\rho_R^\pm$ do not couple to two light fermions at this order.

Inserted in the tree-level dictionary of Ref.~\cite{deBlas:2017xtg}, Eq.~\eqref{eq:gV} gives Eqs.~\eqref{eq:CV3}--\eqref{eq:CVR}.

\subsection{From the symmetric to the physical basis}
\label{app:vphys}

After electroweak symmetry breaking, the Higgs kinetic term mixes the light and heavy vectors. With the rotations of Eq.~\eqref{eq:vrot} and the heavy masses of Eq.~\eqref{eq:mV}, the vector mass terms are
\begin{align}
  \mathcal L_{\rm mass}={}&M_{W^*}^2W_\mu^{*+}W^{*-\mu}
  +M_{\rho_R^\pm}^2\rho_{R\mu}^+\rho_R^{-\mu}
  +\frac{v_H^2}{4}\bigl|gW_\mu^++g_L^*c_LW_\mu^{*+}-g_R^*\rho_{R\mu}^+\bigr|^2
  \notag\\
  &+\frac12\bigl(M_{W^*}^2W_\mu^{*3}W^{*3\mu}+M_{B^*}^2B_\mu^*B^{*\mu}
  +M_{\rho^\perp}^2\rho_\mu^\perp\rho^{\perp\mu}\bigr)
  \notag\\
  &+\frac{v_H^2}{8}\Bigl(gW_\mu^3-g'B_\mu+g_L^*c_LW_\mu^{*3}
  -\frac{g_R^*c_Y}{\sqrt2}B_\mu^*-\frac{g_R^*}{\sqrt2}\rho_\mu^\perp\Bigr)^2.
\end{align}
They define the charged and neutral mass matrices $\mathcal M_C^2(v_H)$ and $\mathcal M_N^2(v_H)$ in the bases $\mathcal V_C=(W^+,W^{*+},\rho_R^+)^T$ and $\mathcal V_N=(W^3,B,W^{*3},B^*,\rho^\perp)^T$. Orthogonal rotations $R_C$ and $R_N$ define the physical fields,
\begin{equation}
  \mathcal V_a=R_a\mathcal V_a^{\rm phys},\qquad
  R_a^T\mathcal M_a^2(v_H)R_a
    =\operatorname{diag}(m_{a,1}^2,\ldots,m_{a,n_a}^2),
  \qquad a=C,N,
\end{equation}
with $\mathcal V_C^{\rm phys}=(W^+,W_1^+,W_2^+)^T$ and $\mathcal V_N^{\rm phys}=(A,Z,Z_1,Z_2,Z_3)^T$. The Higgs coupling matrices $G_h^a=\partial\mathcal M_a^2/\partial v_H$ rotate in the same way, and their $WW$ and $ZZ$ entries give $\kappa_W$ and $\kappa_Z$ in Eq.~\eqref{eq:kV}.

\subsection{Spin-one contributions to the \texorpdfstring{$S$ and $T$}{S and T} parameters}
\label{app:vST}

The spin-one resonances modify the electroweak vacuum polarizations already at tree level. With the conventions of Appendix~\ref{app:vsym}, a direct integration of the neutral resonances gives
\begin{equation}
S_V^{\rm tree}
=
\frac{8\pi v^2}{f^2}
\left(
1-\frac{v^2}{2f^2}
\right)
\left(
\frac{1}{g_L^{*2}}
+
\frac{1}{g_R^{*2}}
\right),
\label{eq:Sv}
\end{equation}
where $v=(\sqrt{2}G_F)^{-1/2}$. Equation~\eqref{eq:Sv} retains the full tree-level dependence on electroweak symmetry breaking within the Lagrangian of Eq.~\eqref{eq:Lv}, with no expansion in the gauge mixing angles. Its leading term in $v^2/f^2$ reproduces the standard result for two-site constructions~\cite{Contino:2006nn}.

The custodial structure of the spin-one sector protects the $T$ parameter at tree level, yielding
\begin{equation}
T_V^{\rm tree}=0,
\label{eq:Tv}
\end{equation}
again in agreement with the standard two-site result \cite{Contino:2006nn}.

We restrict the spin-one sector to this tree-level treatment. One-loop corrections involving the heavy vectors are higher order effects of the resonance effective theory. How calculable they are depends on the symmetry structure of the spin-one sector, and more symmetric two-site models can show additional cancellations at one loop~\cite{Contino:2015mha}. Establishing this for the present $SU(2)_L\times SU(2)_R\times U(1)_X$ construction would require a dedicated NLO calculation, so we do not assume these contributions vanish but leave them outside the accuracy of the present treatment.

Estimates in related constructions suggest that these loops are subleading with respect to the fermionic ones: for $T$, the diagrams with top Yukawa insertions dominate over the purely gauge ones~\cite{Agashe:2003zs}, and for $S$, the one-loop contributions from the UV dynamics are typically smaller than the logarithmically enhanced fermion loops~\cite{Grojean:2013qca}.

\bibliographystyle{JHEP}
\bibliography{references}
\end{document}